\documentclass[preprint,authoryear,12pt]{elsarticle}
\usepackage{graphicx,color}
\usepackage{amssymb}
\usepackage{natbib}

\journal{Computational Physics}

\newcommand {\vect}[1]{\mbox{\boldmath $#1$}}

\newcommand {\inty}[2]{\int_{#1}^{#2}}

\newcommand {\dif}[3][]{\frac{d^{#1}#2}{d#3^{#1}}}
\newcommand {\pdif}[3][]{\frac{\partial^{#1}#2}{\partial#3^{#1}}}

\def\mart{\@ifnextchar[{\mart@@}{\mart@}}
\def\mart@@[#1]#2{\sqrt[#1]{\mathstrut{#2}}}
\def\mart@#1{\sqrt{\mathstrut{#1}}}
\newcommand {\Alfven}{Alfv\'{e}n}

\newcommand{\sgn}{{\rm sgn}}

\long\def\symbolfootnote[#1]#2{\begingroup%
\def\thefootnote{\fnsymbol{footnote}}\footnote[#1]{#2}\endgroup}

\newcommand{\apj}{Astrophysical Journal }
\newcommand{\jgr}{Journal of Geophysical Research }

\begin{document}

\begin{frontmatter}
\title{Multi-Moment Advection scheme for Vlasov simulations}
\author{Takashi Minoshima\corref{cor1}\fnref{label1}}
\ead{minoshim@jamstec.go.jp}
\author{Yosuke Matsumoto\fnref{label2}}
\author{Takanobu Amano\fnref{label3}}
\cortext[cor1]{Corresponding author. Tel.: +81-45-778-5887; Fax: +81-45-778-5490}
\address[label1]{Institute for Research on Earth Evolution, Japan Agency for Marine-Earth Science and Technology, 3173-25, Syowa-machi, Kanazawaku, Yokohama 236-0001, Japan}
\address[label2]{Solar-Terrestrial Environment Laboratory, Nagoya University,
Furo-cho, Chikusa-ku, Nagoya 464-8601, Japan}
\address[label3]{Department of Physics, Nagoya University, Furo-cho, Chikusa-ku, Nagoya 464-8602, Japan}

\begin{abstract}
We present a new numerical scheme for solving the advection equation and its application to Vlasov simulations.
% The scheme independently solves piecewise values of the zeroth to second order moments as well as a profile itself as dependent variables, to conserve the information entropy of the profile with high accuracy.
% The scheme treats cell-integrated values of the zeroth to second order moments as well as point values of a profile itself as dependent variables, for better conservation of the information entropy.
The scheme treats not only point values of a profile but also its zeroth to second order piecewise moments as dependent variables, for better conservation of the information entropy.
We have developed one- and two-dimensional schemes and show that they provide quite accurate solutions within reasonable usage of computational resources compared to other existing schemes. 
The two-dimensional scheme can accurately solve the solid body rotation problem of a gaussian profile for more than hundred rotation periods with little numerical diffusion. This is crucially important for Vlasov simulations of magnetized plasmas. 
Applications of the one- and two-dimensional schemes to electrostatic and electromagnetic Vlasov simulations are presented with some benchmark tests.
\end{abstract}

\begin{keyword}
Advection equation \sep Conservative form \sep Multi-moment \sep Information entropy \sep Vlasov simulations \sep Magnetized plasmas
\end{keyword}

\end{frontmatter}

\section{Introduction}\label{sec:introduction}
The kinematics of collisionless plasmas has been studied in a wide variety of fields, such as in laboratory plasma physics, space physics, and astrophysics. Evolution of collisionless plasmas and self-consistent electromagnetic fields is fully described by the Vlasov-Maxwell (or Vlasov-Poisson) equations. 
Thanks to recent development in computational technology, self-consistent numerical simulations of collisionless plasmas have been successfully performed from the first-principle Vlasov-Maxwell system of equations.

There are two numerical methods to solve the Vlasov equation. The most popular one is the Particle-In-Cell (PIC) method \citep{PIC}, which approximates the plasma by a finite number of macro-particles. Their trajectories calculated from the equation of motion are continuous in space, whereas electromagnetic fields are calculated on grid points in space. 
% The PIC simulation has been widely used for a large variety of plasma phenomena, because a quite robust method for the time advance of particles has been established, the so-called Buneman-Boris method. Because the Buneman-Boris method is symplectic, the PIC conserves the energy as well as mass very well. 
The PIC method has been used for a wide variety of plasma phenomena, because it gives satisfying results even with a relatively small number of particles. 
% The PIC method conserves energy as well as mass very well.
However, the PIC method inherently has the large statistical noise due to an approximation of the distribution function by a finite number of particles. This noise only decreases in $1/\sqrt{N}$ when the number of particles $N$ is increased, making it difficult to study such as particle acceleration and thermal transport processes, in which a small number of high energy particles play an important role.

To overcome this problem, an alternative method free from the statistical noise has been used, in which the Vlasov equation is directly discretized on grid points in phase space. 
% Since the Vlasov equation is a non-diffusive multidimensional advection equation of the distribution function $f(\vect{x},\vect{v})$, the Vlasov simulation solves the advection equation in multidimensions (up to six). However, it is widely known that an accurate solution of the advection equation is not easy to be obtained numerically even in the one dimension.
% Although the Vlasov equation has a form of the advection equation in multidimensions (up to six), it has been widely known that a numerical solution of the advection equation suffers from spurious oscillations and numerical diffusion.
The so-called Vlasov simulation involves solving the advection equation in multidimensions (up to six).
However, it has been widely known that a numerical solution of the advection equation suffers from spurious oscillations and numerical diffusion.
% A highly accurate scheme is required to preserve the characteristics of the Vlasov equation as much as possible. 
A highly accurate scheme is required to preserve characteristics of the Vlasov equation (i.e., the Liouville theorem) as much as possible. 
In contrast to the PIC simulation, no standard scheme for the Vlasov simulation has been established thus far.

\cite{1976JCoPh..22..330C} proposed a splitting scheme, in which the electrostatic Vlasov equation is split into two advection equations in one-dimensional configuration and velocity spaces, and then are alternately advanced. Both equations reduce to a simple form of the linear advection equation. Moreover, their splitting method is equivalent to the second-order symplectic integration so that the conservation of energy is very well.

Following them, many authors have proposed high order advection schemes and applied them to the Vlasov simulation. One of standard time integration methods for the advection equation is the semi-Lagrangian method, which advances a physical variable by interpolating its profile between grid points and then following the characteristics backward in time. For example, \cite{1999JCoPh.149..201S} and \cite{2009CoPhC.180.1730C} employed this method with the cubic B-spline interpolation. \cite{2001JCoPh.172..166F} developed a semi-Lagrangian scheme in a conservative form with an upwind-biased Lagrange polynomial interpolation, called the Positive and Flux Conservative (PFC) scheme. The PFC preserves mass and positivity. \cite{2008EP&S...60..773U} proposed a non-oscillatory type of the PFC scheme.

The above and many popular schemes consider the time integration of a single physical variable. On the other hand, the concept of ``multi-moment'' that treats multiple dependent variables has been proposed. \cite{1991CoPhC..66..219Y} and \cite{1991CoPhC..66..233Y} developed a multi-moment semi-Lagrangian scheme, called the Constrained Interpolation Profile (CIP) scheme. The CIP scheme employs the cubic Hermite interpolation. Difference of the CIP scheme from the others is that it treats not only a profile but also its first derivative in space as dependent variables, and their governing equations are solved as coupled equations.
% Another type of the semi-Lagrangian scheme has been proposed by \cite{1991CoPhC..66..219Y}, called the Constrained Interpolation Profile (CIP) scheme. The CIP employs the cubic Hermite interpolation for a profile. The difference of the CIP from other schemes is that it treats a profile and its first derivative in space as dependent variables, and then independently advances them based on their governing equations. 
The CIP scheme provides comparable or better solutions even with a relatively smaller number of grid points compared to the others, although it requires a higher memory cost to store multiple dependent variables. 
% \cite{1991CoPhC..66..233Y} have developed the CIP in multidimensions.
Due to its high capability, the CIP scheme has been applied not only to the Vlasov simulation \citep{1999CoPhC.120..122N} but also to magnetohydrodynamic simulations \citep{1998ApJ...508..186K,2008CoPhC.179..289M,2009CoPhC.180.1550Y}.

Conservative schemes are preferable for the Vlasov simulation, because the integration of the distribution function in phase space is equal to the number of particles.
\cite{2001mwr...129..332Y} and \cite{2002CoPhC.148..137T} also proposed a conservative form of the CIP, called the CIP-CSL2 (CIP-Conservative Semi-Lagrangian scheme with a second-order polynomial). The CIP-CSL2 scheme treats point values of a profile and its cell-integrated values as dependent variables. 
% The profile in a cell is interpolated by using the variables at the upwind position as constraints.
The cell-integrated value is advanced in a conservative form to guarantee the conservation of mass. 
Therefore, the CIP-CSL2 scheme will be more suitable than the CIP scheme for the Vlasov simulation. 
{ Various CIP-CSL type schemes have been proposed thus far \citep[e.g.,][]{2000CoPhC.126..232T}.
% For example, \cite{2000CoPhC.126..232T} and \cite{2009JCoPh.228.3669I} have developed higher-order schemes, by increasing the number of dependent variables to be treated.
% \cite{2000CoPhC.126..232T} developed a higher order scheme, which treats point values of a profile and its first derivative, and cell-integrated values as dependent variables.
% \cite{2005CoPhC.173...17I} proposed a two-dimensional scheme on triangular unstructured grids.
% \cite{2009JCoPh.228.3669I} developed up to sixth-order accurate schemes, by increasing the number of dependent variables within a cell.
A non-oscillatory type of the CIP-CSL scheme was proposed by \cite{2001JCoPh.170..498X}, and is successfully extended to fluid dynamic equations \citep{2004JCoPh.195..629X,2006JCoPh.213...31X,2007JCoPh.222..849I}.
\cite{2008JCoPh.227.2263I} developed a conservative Eulerian scheme, which is applied to the electrostatic Vlasov-Poisson simulation \citep{2009JCoPh.228.8919I}. }

% The multidimensional CIP-CSL2 has been developed by \cite{2002CoPhC.148..137T}.
% \cite{2000CoPhC.126..232T} have developed a higher order scheme called CIP-CSL4, in which a profile, its first derivative in space, and piecewise integrated value in a cell are treated as dependent variables.
% An IDO type form of the CIP-CSL2 has been developed by \cite{2008JCoPh.227.2263I}, called IDO-CF (Conservative Form), and been applied to the Vlasov equation \citep{2009JCoPh.228.8919I}.

% The concept of the CIP schemes that treat multiple dependent variables and advance them based on their governing equations is called ``multi-moment'' \citep{2004JCoPh.195..629X}. Another type of the multi-moment scheme with the Eulerian time integration has been proposed by \cite{1997CoPhC.102..132A}, called the Interpolated Differential Operator (IDO) scheme. The IDO scheme in a conservative form has been developed by \cite{2008JCoPh.227.2263I}, called IDO-CF, and been applied to the Vlasov equation by \cite{2009JCoPh.228.8919I}.

% However, most of previously-proposed schemes are limited to be applied to the one-dimensional Vlasov-Poisson simulation or the guiding-center (drift/gyro-kinetic) simulation, in which the gyro motion of particles around the magnetic field is not solved. The electromagnetic Vlasov simulation is rare \citep[e.g.,][]{2002JCoPh.179..495M,2006PhPl...13i2309S,2009CoPhC.180..365U,2010PhPl...17e2311U}.
Although there has been a number of numerical schemes, they have been applied mainly to the electrostatic Vlasov-Poisson simulation.
There is an increasing interest for the application of electromagnetic Vlasov simulations to magnetized plasmas \citep{2002JCoPh.179..495M,2006PhPl...13i2309S,2009JCoPh.228.4773S,2010JCoPh.229.1643S,2009CoPhC.180..365U,2010PhPl...17e2311U}.
However, the Vlasov simulation of magnetized plasmas is more difficult than the electrostatic one, because no suitable scheme for solving the gyro motion (solid body rotation in velocity space) has been established thus far.
% There are a few reports that attempt to apply them to the electromagnetic Vlasov-Maxwell equation system \citep{2002JCoPh.179..495M,2006PhPl...13i2309S,2009CoPhC.180..365U,2010PhPl...17e2311U,2009JCoPh.228.4773S,2010JCoPh.229.1643S}.
% This is mostly due to the difficulty in numerically solving the gyro motion of a magnetized plasma (solid body rotation in velocity space). 
% An advection scheme with insufficient accuracy causes serious numerical diffusion in the solid body rotation problem.
\cite{2006CoPhC.175...86S} proposed a rigorous time-splitting method to solve the solid body rotation problem using the one-dimensional PFC scheme, called the backsubstitution. 
% Even with this method, however, one can not avoid considerable numerical diffusion in a long time calculation. 
Even with this method, however, we find considerable numerical diffusion in a long time calculation that causes non-physical plasma heating during the gyration. 
% Even with this method or the multidimensional CIP schemes, however, it is almost impossible to solve the problem for a long time without numerical diffusion. 
% A physically-meaningful electromagnetic Vlasov simulation is still difficult to be performed.
For this reason, it is still quite limited to deal with plasma phenomena such as heating and acceleration on the basis of the electromagnetic Vlasov simulation.
% To overcome this problem, it is necessary to develop an extremely accurate scheme for the advection equation in multidimensions.

In this paper, we propose a new numerical scheme for the advection equation, specifically designed to solve the Vlasov equation in magnetized plasmas. We first argue in Section \ref{sec:concept-mma-scheme} that it is important to preserve high order moments of the distribution function to reduce numerical diffusion, on the basis of the concept of the conservation of the information entropy.
%  We develop an advection scheme, named as ``Multi-Moment Advection (MMA)'', in which piecewise values of the zeroth to second order moments as well as a profile itself are treated as dependent variables.
% Details of the MMA are described in Sections \ref{sec:mma-one-dimension} and \ref{sec:mma-two-dimension} for one and two dimensions, respectively. Benchmark tests of the MMA and its application to the electrostatic and electromagnetic Vlasov simulations are presented in Sections \ref{sec:numer-tests-mma1d} and \ref{sec:numer-tests-mma2d}. Finally we summarize the paper in Section \ref{sec:conclusion}. 
 We develop an advection scheme, in which not only point values of a profile but also its zeroth to second order piecewise moments are treated as dependent variables.
Details of the scheme are described in Sections \ref{sec:mma-one-dimension} and \ref{sec:mma-two-dimension} for one and two dimensions, respectively. Benchmark tests of the scheme and its application to electrostatic and electromagnetic Vlasov simulations are presented in Sections \ref{sec:numer-tests-mma1d} and \ref{sec:numer-tests-mma2d}. Finally, we summarize the paper in Section \ref{sec:conclusion}. 
%  Details of the one dimensional MMA (MMA1D) are described in Section \ref{sec:mma-one-dimension}. Test simulations of the MMA1D and its application to the electrostatic Vlasov simulation are shown in Section \ref{sec:numer-tests-mma1d}. Details of the two dimensional MMA (MMA2D) are described in Section \ref{sec:mma-two-dimension}. Test simulations of the MMA2D and its application to the electromagnetic Vlasov simulation are shown in Section \ref{sec:numer-tests-mma2d}. Finally we conclude the paper in Section \ref{sec:conclusion}. 

\section{Concept of Multi-Moment Advection scheme}\label{sec:concept-mma-scheme}
{ We consider that the conservation of the information entropy is essential to develop a dissipationless scheme for the advection equation. 
When a profile $f(x)$ follows the advection equation, $Df/Dt = 0$, its entropy function, $h(x) = f(x) I(f(x))$, also follows the equation,
\begin{eqnarray}
\frac{D h}{D t} = \left(f \dif{I}{f} + I \right) \frac{D f}{D t} = 0,\nonumber 
\end{eqnarray}
where $I(f(x))$ is the self-information function of $x$, and $f(x)$ is regarded as the probability function that rapidly decreases toward $x=\pm \infty$. The conventional information entropy function is $h = -f \log f$. \cite{1988JSP....52..479T} defined more general function of the form, $h = \left(f-f^q\right)/\left(q-1\right)$. 
The shape of a profile will be preserved during the advection with constant velocity when its information entropy is conserved in a numerical simulation,
\begin{eqnarray}
H=\inty{-\infty}{\infty}h(x)dx = \sum_{i}^{}\inty{x_i}{x_{i+1}} f(x) I(f(x)) dx = {\rm Constant},\nonumber
\end{eqnarray}
where $x_i$ is the position in physical space.}
% We develop a high order scheme for the linear advection equation from the viewpoint of the conservation of the information entropy.
% We consider that the conservation of the information entropy is essential to develop a dissipationless scheme for the linear advection equation. 
% The shape of a profile $f(x)$ will be preserved during the linear advection when its information entropy,
% % It is expected that the shape of a profile $f(x)$ will be exactly kept during the linear advection, if its information entropy is exactly conserved, that is,
% \begin{eqnarray}
% H=-\sum_{i}^{}\inty{x_i}{x_{i+1}} f(x) \log f(x)dx,\nonumber
% \end{eqnarray}
% or more generally,
% \begin{eqnarray}
% H=\sum_{i}^{}\inty{x_i}{x_{i+1}} f(x) I(f(x))dx,\nonumber
% \end{eqnarray}
% is conserved in a numerical simulation. Here $x_i$ is the position in the physical space, $I(f(x))$ is the self-information function of $x$, and $f(x)$ is regarded as the probability function that rapidly decreases toward $x=\pm \infty$.
However, it is difficult to develop an advection scheme that exactly guarantees the entropy conservation, because $I(f(x))$ is nonlinear in general. 
% However, we can easily rewrite it as a linear combination by the Taylor expansion with respect to $x$,
Then, we approximate it by expanding in the Taylor series with respect to $x$,
% \begin{eqnarray}
% H &=& \inty{}{} f\left[I\left(f(x_0)\right) + \left.\pdif{I}{x}\right|_{x_0}\left(x-x_0\right) + \left.\frac{1}{2}\frac{\partial^2 I}{\partial x^2}\right|_{x_0}\left(x-x_0\right)^2 + \dots \right]dx \nonumber \\
% &=& \left(I-\pdif{I}{x} x_{0}+\frac{1}{2}\frac{\partial^2 I}{\partial x^2} x_0^2-\dots\right)_{x_0} \inty{}{}fdx\nonumber \\
% && + \left(\pdif{I}{x} - \frac{\partial^2 I}{\partial x^2} x_0 +\dots\right)_{x_0} \inty{}{} xfdx \nonumber \\
% && + \left(\frac{\partial^2 I}{\partial x^2} - \dots\right)_{x_0} \inty{}{} \frac{x^2}{2}fdx + \dots.\label{eq:74}
% \end{eqnarray}
\begin{eqnarray}
H &=& \sum_{i}^{}\inty{x_i}{x_{i+1}} f\left[I(f(0)) + \left.\pdif{I}{x}\right|_{x=0}x + \left.\frac{1}{2}\frac{\partial^2 I}{\partial x^2}\right|_{x=0}x^2 + \dots \right]dx.\nonumber
\end{eqnarray}
This equation tells that the entropy can be described as a linear combination of the zeroth to $n$-th order moments, $\inty{}{}(x^{n}f/n!)dx$. We then consider that better conservation of the entropy may be achieved by preserving information of high order moments. 
% For example, the entropy of a gaussian profile, which is the most fundamental distribution in the particle velocity space and hence we are the most interested in, is completely described with a combination up to the second order moment.
% For example, the entropy of a one-dimensional gaussian profile, $f(x) = \exp[-(x-x_0)^2/2\sigma_x^2]/\sqrt{2 \pi}\sigma_x$, which is the most fundamental distribution in the velocity space and maximizes the entropy in the thermal equilibrium, is described by up to the second order moment as
For example, the entropy of a gaussian profile, $f(x) = \exp[-(x-x_0)^2/2\sigma_x^2]/\sqrt{2 \pi}\sigma_x$, is described by up to the second order moment,
\begin{eqnarray}
H &=& -\sum_{i}^{}\inty{x_i}{x_{i+1}} f(x) \log f(x)dx\nonumber \\
&\propto& \sum_{i}^{}\inty{x_i}{x_{i+1}}f(x) \left(x-x_0\right)^2 dx \nonumber\\
&=& \sum_{i}^{}\left[x_0^2\inty{x_{i}}{x_{i+1}}fdx-2x_0\inty{x_{i}}{x_{i+1}}xfdx + \inty{x_{i}}{x_{i+1}}x^2fdx\right].\nonumber
\end{eqnarray} 
Since the gaussian profile is the most fundamental distribution in velocity space that maximizes the entropy in thermal equilibrium, preserving the entropy of this profile is crucially important for Vlasov simulations.
We speculate from this discussion that it is essential to treat, at least, up to the second order moment as dependent variables, to keep the characteristics of the Vlasov equation. A ``Multi-Moment Advection (MMA)'' scheme is designed to treat the zeroth to second order moments as well as the profile itself as dependent variables, and advance them on the basis of their governing equations.
% Note that although the term of ``multi-moment'' has been already introduced by \cite{2004JCoPh.195..629X}, our concept is slightly different from it in that the moment in our scheme has a meaning in statistics.

\section{One-dimensional MMA scheme (MMA1D)}\label{sec:mma-one-dimension}
We consider the time evolution of a one-dimensional profile $f(x,t)$ and its zeroth to second order moments defined as
\begin{eqnarray}
M^m = \frac{1}{m!}\inty{}{}x^{m}fdx,\;\;\; \left(m=0,1,2\right).\label{eq:4}
\end{eqnarray}
% Their advection equations are written in the following conservative form,
% \begin{eqnarray}
% && \pdif{f}{t} + \pdif{}{x}\left(uf\right) = 0,\label{eq:1}\\
% && \pdif{M^0}{t} + u\pdif{M^0}{x} = 0,\label{eq:2}\\
% && \pdif{M^m}{t} + u\pdif{M^m}{x} = \frac{1}{\left(m-1\right)!}\inty{}{}ux^{m-1}fdx,\;\;\; \left(m=1,2\right),\label{eq:3}
% \end{eqnarray}
% where $u$ is the velocity.
{ The conservative advection equation of $f$ and governing equations of the moments are written as
\begin{eqnarray}
&& \pdif{f}{t} + u\pdif{f}{x} = -\pdif{u}{x}f,\label{eq:1}\\
&& \pdif{M^0}{t} + \inty{}{}dx \pdif{}{x}\left(u f\right) = 0,\label{eq:2}\\
&& \pdif{M^m}{t} + \frac{1}{m!} \inty{}{}dx \pdif{}{x}\left(u x^{m} f\right)= \frac{1}{\left(m-1\right)!}\inty{}{}ux^{m-1}fdx,\;\;\; \left(m=1,2\right),\label{eq:3}
\end{eqnarray}
where $u$ is the velocity. Here, the conservative advection equation of $f$ is cast into the advective form.}
% { The advection equation of $f$ and governing equations of the moments are written as
% \begin{eqnarray}
% && \pdif{f}{t} + u\pdif{f}{x} = -\pdif{u}{x}f,\label{eq:1}\\
% && \pdif{M^0}{t} + \inty{}{}d \left(u f\right) = 0,\label{eq:2}\\
% && \pdif{M^m}{t} + \frac{1}{m!} \inty{}{}d \left(u x^{m} f\right)= \frac{1}{\left(m-1\right)!}\inty{}{}ux^{m-1}fdx,\;\;\; \left(m=1,2\right),\label{eq:3}
% \end{eqnarray}
% where $u$ is the velocity.}
Equations (\ref{eq:2}) and (\ref{eq:3}) are obtained by multiplying Equation (\ref{eq:1}) by $x^{m}/m!$ and then integrating over space.
Note that the CIP-CSL2 scheme considers Equations (\ref{eq:1}) and (\ref{eq:2}) ($m=0$ only).
% Equations (\ref{eq:3}) as well as (\ref{eq:2}) are exactly derived by integrating Equation (\ref{eq:1}) with $x$.
Hereafter, we assume the constant velocity case, $\partial u / \partial x = 0$, because we are concerned with the Vlasov equation.

To solve a set of these equations, the one-dimensional MMA scheme (MMA1D) treats four dependent variables; the point value of the profile $f_i$, and the piecewise moments,
\begin{eqnarray}
M^m_{i+1/2} = \frac{1}{m!}\inty{x_{i}}{x_{i+1}}x^{m}fdx,\;\;\; \left(m=0,1,2\right),\label{eq:5}
\end{eqnarray}
and constructs a piecewise interpolation for $f$ in a cell with a fourth order polynomial,
\begin{eqnarray}
F_{i}(x) &=& \sum_{k=1}^{5} k C_{k;i} \left(x-x_i\right)^{k-1},\label{eq:6}
\end{eqnarray}
which gives an interpolation function for $M^m$ as
\begin{eqnarray}
G^m_{i}(x) &=& \frac{1}{m!} \inty{x_{i}}{x}x^m F_{i}(x)dx \nonumber \\
&=& \sum_{k=1}^{5} A^m_{k}(x,x_{i}) C_{k;i} \left(x-x_i\right)^{k},\;\;\; \left(m=0,1,2\right),\label{eq:7}
\end{eqnarray}
where
\begin{eqnarray}
\left\{
\begin{array}{lll}
A^0_{k}(x,x_{i}) &=& 1,\\
A^1_{k}(x,x_{i}) &=& \left(kx+x_i\right) / \left(k+1\right),\\
A^2_{k}(x,x_{i}) &=& \left\{\frac{k\left(k+1\right)}{2}x^2 + kx_{i}x + x_{i}^2 \right\} / \left\{\left(k+1\right)\left(k+2\right)\right\}.\label{eq:10}
\end{array}
\right.
\end{eqnarray}

To determine the coefficients $C_{k;i}$, we use the variables at the upwind position as constraints,
\begin{eqnarray}
\left\{
\begin{array}{lll}
F_{i}(x_{i}) &=& f_{i},\\
F_{i}(x_{iup}) &=& f_{iup},\\
G^m_{i}(x_{iup}) &=& \sgn \left(\zeta \right) M^m_{icell}, \;\;\; (m=0,1,2), \label{eq:13}
\end{array}
\right.
\end{eqnarray}
where
\begin{eqnarray}
\left\{
\begin{array}{lll}
iup &=& i + \sgn \left(\zeta\right),\\
icell &=& i + \sgn \left(\zeta\right)/2,\label{eq:15}
\end{array}
\right.
\end{eqnarray}
are the position of the upwind grid and cell, $\sgn(\zeta)$ stands for the sign of $\zeta$, and $\zeta = -u\Delta t$ is the distance of the upwind departure position relative to $x_{i}$.
Consequently, the coefficients are explicitly determined as
{\normalsize
\begin{eqnarray}
C_{1;i} &=& f_{i},\label{eq:17}\\
C_{2;i} &=& \frac{-2}{\Delta x} \left[ 4f_{i} + f_{iup} - \frac{15 \sgn \left(\zeta\right)}{\Delta x^3} \times \right. \nonumber \\
&& \left. \left\{ \left(2x_{iup}^2 + 4x_{i}x_{iup} + x_{i}^2 \right)M^0_{icell} \right.\right. \nonumber \\
&& \left.\left. - 2\left(4x_{iup}+3x_{i}\right) M^1_{icell} + 14 M^2_{icell} \right\}\right], \label{eq:18}\\
C_{3;i} &=& \frac{10}{\Delta x^2} \left[ 2f_{i} + f_{iup} - \frac{2 \sgn \left(\zeta\right)}{\Delta x^3} \times \right. \nonumber \\
&& \left. \left\{ \left(10x_{iup}^2 + 25x_{i}x_{iup} + 7x_{i}^2 \right)M^0_{icell}\right.\right.\nonumber \\
&& \left.\left. - \left(45x_{iup}+39x_{i}\right) M^1_{icell} + 84 M^2_{icell} \right\}\right], \label{eq:19}\\
C_{4;i} &=& \frac{-5}{\Delta x^3} \left[ 4f_{i} + 3f_{iup} - \frac{3 \sgn \left(\zeta\right)}{\Delta x^3} \times \right. \nonumber \\
&& \left. \left\{ \left(15x_{iup}^2 + 42x_{i}x_{iup} + 13x_{i}^2 \right)M^0_{icell}\right.\right.\nonumber \\
&& \left.\left. - 4\left(18x_{iup}+17x_{i}\right) M^1_{icell} + 140 M^2_{icell} \right\}\right], \label{eq:20}\\
C_{5;i} &=& \frac{7}{\Delta x^4} \left[ f_{i} + f_{iup} - \frac{12 \sgn \left(\zeta\right)}{\Delta x^3} \times \right. \nonumber \\
&& \left. \left\{ \left(x_{iup}^2 + 3x_{i}x_{iup} + x_{i}^2 \right)M^0_{icell}\right.\right.\nonumber\\
&& \left.\left. - 5\left(x_{iup}+x_{i}\right) M^1_{icell} + 10 M^2_{icell} \right\}\right]. \label{eq:21}
\end{eqnarray}
}where $\Delta x = x_{iup}-x_i$.

Then, let us consider the time integration of the variables. The MMA1D employs the semi-Lagrangian method. Equations (\ref{eq:1})-(\ref{eq:3}) are split into two phases, advection and non-advection phases. We first calculate the advection phase as
{\footnotesize
\begin{eqnarray}
{}^{n+1}f_{i} &=& {}^{n}F_{i} (x_{i} + \zeta),\label{eq:22}\\
{}^{n+1}M^0_{i+1/2} &=& {}^{n}M^0_{i+1/2} + \inty{x_{i+1}}{x_{i+1}+\zeta} {}^{n}F_{i+1}(x)dx - \inty{x_{i}}{x_{i}+\zeta} {}^{n}F_{i}(x)dx \nonumber \\ 
&=& {}^{n}M^0_{i+1/2} + \left\{{}^nG^0_{i+1}(x_{i+1}+\zeta) -  {}^nG^0_{i}(x_{i}+\zeta) \right\}, \label{eq:23}\\
{}^*M^m_{i+1/2} &=& {}^nM^m_{i+1/2} + \frac{1}{m!}\left[\inty{x_{i+1}}{x_{i+1}+\zeta} x^{m} \cdot {}^{n}F_{i+1}(x)dx - \inty{x_{i}}{x_{i}+\zeta} x^{m} \cdot {}^{n}F_{i}(x)dx\right] \nonumber \\
&=& {}^nM^m_{i+1/2} + \left\{{}^nG^m_{i+1}(x_{i+1}+\zeta) - {}^nG^m_{i}(x_{i}+\zeta) \right\},\; \left(m=1,2\right),\label{eq:24}
\end{eqnarray}
}where the left-superscript indicates the number of time steps and the asterisk means that the variables are at the intermediate step. Here, we consider the case of the CFL number $|u \Delta t / \Delta x| < 1$ for simplicity. 
% Obviously, Equation (\ref{eq:23}) guarantees the conservation of mass. 
We next advance the non-advection phase. Since $u$ is constant, the right-hand side of Equation (\ref{eq:3}) is equal to $u M^{m-1}$.
% \begin{eqnarray}
% \frac{1}{\left(m-1\right)!}\inty{x_{i}}{x_{i+1}} ux^{m-1}fdx = u M^{m-1}_{i+1/2},\;\;\; \left(m=1,2\right).\nonumber
% \end{eqnarray}
Therefore, the non-advection phase is calculated as
 \begin{eqnarray}
% {}^{n+1}f_{i} &=& {}^*f_{i} \left\{ 1 - \left(\pdif{u}{x}\right)_{i} \Delta t \right\},\label{eq:26}\\
{}^{n+1}M^1_{i+1/2} &=& {}^*M^1_{i+1/2} + u{}^{n+1}M^0_{i+1/2} \Delta t,\label{eq:27}\\
{}^{n+1}M^2_{i+1/2} &=& {}^*M^2_{i+1/2} + u \left({}^*M^1_{i+1/2} + u {}^{n+1}M^0_{i+1/2} \frac{\Delta t}{2} \right) \Delta t.\label{eq:28}
\end{eqnarray}

% Similar to the CIP-CSL2, the MMA1D with the semi-Lagrangian method can be applied to a large CFL number by simply considering the cell in Equation (\ref{eq:7}) to be the cell at the upwind departure position, and then replacing $i$ in the right-hand side of Equations (\ref{eq:22})-(\ref{eq:24}) by $i-{\rm Int}(u\Delta t/|\Delta x|)$, where ${\rm Int}(x)$ returns the integer number of $x$ and a uniform grid spacing is assumed.
{
It is noteworthy that in the advection problem with constant velocity, the MMA scheme exactly guarantees the conservation of the zeroth to second order central moments defined as
\begin{eqnarray}
\frac{1}{m!} \inty{-\infty}{\infty} \left(x-u t\right)^m f dx, \;\;\; (m=0,1,2).\label{eq:25} 
\end{eqnarray}
The conservation of the zeroth order central moment (i.e., mass) is evidently guaranteed by Equation (\ref{eq:23}). Using Equations (\ref{eq:24}) through (\ref{eq:28}), the conservation of the first order central moment is obtained as
% \begin{eqnarray}
% && \sum_{i} \left({}^{n+1}M^1_{i+1/2} - u \Delta t {}^{n+1}M^0_{i+1/2}\right) = \sum_{i} {}^*M^1_{i+1/2} = \sum_{i} {}^{n}M^1_{i+1/2} \nonumber \\
% &\Rightarrow& \inty{-\infty}{\infty} \left(x-u \Delta t\right) \cdot {}^{n+1}f dx = \inty{-\infty}{\infty} x \cdot {}^{n}f dx,
% \end{eqnarray}
\begin{eqnarray}
&& \inty{-\infty}{\infty} \left(x-u \Delta t\right) \cdot {}^{n+1}f dx \nonumber \\
&=& \sum_{i} \left({}^{n+1}M^1_{i+1/2} - u \Delta t {}^{n+1}M^0_{i+1/2}\right) = \sum_{i} {}^*M^1_{i+1/2} \nonumber \\
&=& \sum_{i} {}^{n}M^1_{i+1/2} = \inty{-\infty}{\infty} x \cdot {}^{n}f dx,
\end{eqnarray}
and that of the second order central moment (i.e., variance) is also as
% \begin{eqnarray}
% \sum_{i} {}^{n+1}M^2_{i+1/2} = \sum_{i} {}^*M^2_{i+1/2} + u \Delta t \sum_{i} \left({}^*M^1_{i+1/2} + u {}^{n+1}M^0_{i+1/2} \frac{\Delta t}{2} \right) \nonumber \\
% = \sum_{i} {}^{n}M^2_{i+1/2} + u \Delta t \sum_{i} {}^{n}M^1_{i+1/2} + \frac{\left(u \Delta t\right)^2}{2} \sum_{i} {}^{n+1}M^0_{i+1/2} \nonumber \\
% = \sum_{i} {}^{n}M^2_{i+1/2} + u \Delta t \sum_{i} {}^{n+1}M^1_{i+1/2} - \frac{\left(u \Delta t\right)^2}{2} \sum_{i} {}^{n+1}M^0_{i+1/2} \nonumber \\
% \sum_{i} \left({}^{n+1}M^2_{i+1/2} - u \Delta t {}^{n+1}M^1_{i+1/2} + \frac{\left(u \Delta t\right)^2}{2} {}^{n+1}M^0_{i+1/2} \right) = \sum_{i} {}^{n}M^2_{i+1/2} \nonumber \\
% \frac{1}{2} \inty{-\infty}{\infty} \left(x-u \Delta t\right)^2 \cdot {}^{n+1}f dx = \frac{1}{2} \inty{-\infty}{\infty} x^2 \cdot {}^{n}f dx.
% \end{eqnarray}
% \begin{eqnarray}
% && \sum_{i} \left({}^{n+1}M^2_{i+1/2} - u \Delta t {}^{n+1}M^1_{i+1/2} + \frac{\left(u \Delta t\right)^2}{2} {}^{n+1}M^0_{i+1/2} \right) \nonumber \\
% && = \sum_{i} \left[{}^{n}M^2_{i+1/2} + u \Delta t \left({}^*M^1_{i+1/2} - {}^{n+1}M^1_{i+1/2} + u \Delta t {}^{n+1}M^0_{i+1/2} \right) \right]\nonumber \\
% && = \sum_{i} {}^{n}M^2_{i+1/2} \nonumber \\
% &\Rightarrow& \frac{1}{2} \inty{-\infty}{\infty} \left(x-u \Delta t\right)^2 \cdot {}^{n+1}f dx = \frac{1}{2} \inty{-\infty}{\infty} x^2 \cdot {}^{n}f dx.
% \end{eqnarray}
\begin{eqnarray}
&& \frac{1}{2} \inty{-\infty}{\infty} \left(x-u \Delta t\right)^2 \cdot {}^{n+1}f dx \nonumber \\
&=& \sum_{i} \left({}^{n+1}M^2_{i+1/2} - u \Delta t {}^{n+1}M^1_{i+1/2} + \frac{\left(u \Delta t\right)^2}{2} {}^{n+1}M^0_{i+1/2} \right) \nonumber \\
&=& \sum_{i} \left[{}^{*}M^2_{i+1/2} + u \Delta t \left({}^*M^1_{i+1/2} - {}^{n+1}M^1_{i+1/2} + u \Delta t {}^{n+1}M^0_{i+1/2} \right) \right]\nonumber \\
&=& \sum_{i} {}^{n}M^2_{i+1/2} = \frac{1}{2} \inty{-\infty}{\infty} x^2 \cdot {}^{n}f dx.
\end{eqnarray}
}
\section{Two-dimensional MMA scheme (MMA2D)}\label{sec:mma-two-dimension}
We consider the time evolution of a two-dimensional profile $f(x,y,t)$ and its zeroth to second order moments in the $x$ and $y$ directions defined as
\begin{eqnarray}
M^0 &=& \inty{}{} \!\!\! \inty{}{} fdxdy \left(=M_x^0=M_y^0\right),\label{eq:11}\\
M^m_{x} &=& \frac{1}{m!} \inty{}{} \!\!\! \inty{}{} x^m fdxdy, \;\;\; \left(m=1,2\right),\label{eq:8}\\
M^m_{y} &=& \frac{1}{m!} \inty{}{} \!\!\! \inty{}{} y^m fdxdy, \;\;\; \left(m=1,2\right). \label{eq:9}
\end{eqnarray}
% The advection equation of $f$ and governing equations of the moments are written in the following conservative form,
% \begin{eqnarray}
% && \pdif{f}{t} + \pdif{}{x} \left(uf\right) + \pdif{}{y} \left(vf\right) = 0,\label{eq:12}\\
% && \pdif{M^0}{t} + \inty{}{}dx\pdif{}{x} \left(\inty{}{}ufdy\right) + \inty{}{}dy \pdif{}{y} \left(\inty{}{} vfdx\right) = 0,\label{eq:29}\\
% && \pdif{M^m_{x}}{t} + \inty{}{}dx\pdif{}{x} \left(\frac{x^m}{m!} \! \inty{}{}ufdy\right) + \inty{}{}dy \pdif{}{y} \left(\frac{1}{m!} \inty{}{} v x^m fdx\right) \nonumber \\
% && = \frac{1}{\left(m-1\right)!} \inty{}{} \!\!\! \inty{}{} u x^{m-1} fdxdy,\;\;\; \left(m=1,2\right),\label{eq:30}\\
% && \pdif{M^m_{y}}{t} + \inty{}{}dx\pdif{}{x} \left(\frac{1}{m!} \inty{}{}u y^m fdy\right) + \inty{}{}dy \pdif{}{y} \left(\frac{y^m}{m!} \! \inty{}{} vfdx\right)\nonumber \\
% && = \frac{1}{\left(m-1\right)!} \inty{}{} \!\!\! \inty{}{} v y^{m-1} fdxdy,\;\;\; \left(m=1,2\right),\label{eq:31}
% \end{eqnarray}
% where $u$ and $v$ are the velocity components in the $x$ and $y$ directions.
{ The conservative advection equation of $f$ and governing equations of the moments are written as
\begin{eqnarray}
&& \pdif{f}{t} + u \pdif{f}{x} + v \pdif{f}{y} = -\left(\pdif{u}{x}+\pdif{v}{y}\right)f,\label{eq:12}\\
&& \pdif{M^0}{t} + \inty{}{}dx\pdif{}{x} \left(\inty{}{}ufdy\right) + \inty{}{}dy \pdif{}{y} \left(\inty{}{} vfdx\right) = 0,\label{eq:29}\\
&& \pdif{M^m_{x}}{t} + \inty{}{}dx\pdif{}{x} \left(\frac{x^m}{m!} \! \inty{}{}ufdy\right) + \inty{}{}dy \pdif{}{y} \left(\frac{1}{m!} \inty{}{} v x^m fdx\right) \nonumber \\
&& = \frac{1}{\left(m-1\right)!} \inty{}{} \!\!\! \inty{}{} u x^{m-1} fdxdy,\;\;\; \left(m=1,2\right),\label{eq:30}\\
&& \pdif{M^m_{y}}{t} + \inty{}{}dx\pdif{}{x} \left(\frac{1}{m!} \inty{}{}u y^m fdy\right) + \inty{}{}dy \pdif{}{y} \left(\frac{y^m}{m!} \! \inty{}{} vfdx\right)\nonumber \\
&& = \frac{1}{\left(m-1\right)!} \inty{}{} \!\!\! \inty{}{} v y^{m-1} fdxdy,\;\;\; \left(m=1,2\right),\label{eq:31}
\end{eqnarray}
where $u$ and $v$ are the velocity component in the $x$ and $y$ directions.
Here, the conservative advection equation of $f$ is cast into the advective form.}
Equations (\ref{eq:29})-(\ref{eq:31}) are obtained by multiplying Equation (\ref{eq:12}) by $x^{m}/m!$ or $y^{m}/m!$ and then integrating over space.
Hereafter, we assume $\partial u / \partial x = \partial v / \partial y = 0$, but not necessarily $\partial u / \partial y = \partial v / \partial x = 0$, and use vector forms $\vect{M}^{m} = (M_{x}^{m},M_{y}^{m})$, $\vect{x} = (x,y)$, and $\vect{u} = (u,v)$.

 To solve a set of these equations, the two-dimensional MMA scheme (MMA2D) treats six dependent variables; the point value of the profile $f_{i,j}$, and the piecewise moments,
\begin{eqnarray}
\vect{M}^m_{i+1/2,j+1/2} &=& \frac{1}{m!} \inty{y_{j}}{y_{j+1}} \!\!\! \inty{x_{i}}{x_{i+1}} \vect{x}^m fdxdy, \;\;\; \left(m=0,1,2\right),\label{eq:34}
\end{eqnarray}
% \begin{eqnarray}
% M^m_{x;i+1/2,j+1/2} &=& \frac{1}{m!} \inty{y_{j}}{y_{j+1}} \!\!\! \inty{x_{i}}{x_{i+1}} x^m fdxdy,\label{eq:33} \\
% M^m_{y;i+1/2,j+1/2} &=& \frac{1}{m!} \inty{y_{j}}{y_{j+1}} \!\!\! \inty{x_{i}}{x_{i+1}} y^m fdxdy, \;\;\; \left(m=0,1,2\right),\label{eq:34}
% \end{eqnarray}
% \begin{eqnarray}
% M^0_{i+1/2,j+1/2} &=& \inty{y_{j}}{y_{j+1}} \!\!\! \inty{x_{i}}{x_{i+1}} fdxdy,\label{eq:32} \\
% M^m_{x;i+1/2,j+1/2} &=& \frac{1}{m!} \inty{y_{j}}{y_{j+1}} \!\!\! \inty{x_{i}}{x_{i+1}} x^m fdxdy,\label{eq:33} \\
% M^m_{y;i+1/2,j+1/2} &=& \frac{1}{m!} \inty{y_{j}}{y_{j+1}} \!\!\! \inty{x_{i}}{x_{i+1}} y^m fdxdy, \;\;\; (m=1,2),\label{eq:34}
% \end{eqnarray}
and constructs a piecewise interpolation for $f$ in a cell with a polynomial,
\begin{eqnarray}
F_{i,j}(x,y) &=& \sum_{l=1}^{3} \sum_{k=1}^{3} lk C_{lk;i,j} \left(x-x_{i}\right)^{k-1} \left(y-y_{j}\right)^{l-1},\label{eq:35}
\end{eqnarray}
which gives an interpolation function for $\vect{M}^m$ as
{\small
\begin{eqnarray}
\vect{G}^m_{i,j}(x,y) &=& \frac{1}{m!} \inty{y_{j}}{y} \!\!\! \inty{x_{i}}{x} \vect{x}^m F_{i,j}(x,y)dxdy \nonumber \\
&=& \sum_{l=1}^{3} \sum_{k=1}^{3}
\left(
\begin{array}{c}
A^m_{k}(x,x_{i})\\
A^m_{l}(y,y_{j})\\
\end{array}
\right)
C_{lk;i,j} \left(x-x_{i}\right)^{k} \left(y-y_{j}\right)^{l},\label{eq:36}
% G^m_{x;i,j}(x,y) &=& \frac{1}{m!} \inty{y_{j}}{y} \!\!\! \inty{x_{i}}{x} x^m F_{i,j}(x,y)dxdy \nonumber \\
% &=& \sum_{l=1}^{3} \sum_{k=1}^{3} A^m_{k}(x,x_{i})C_{lk;i,j} \left(x-x_{i}\right)^{k} \left(y-y_{j}\right)^{l},\label{eq:36}\\
% G^m_{y;i,j}(x,y) &=& \frac{1}{m!} \inty{y_{j}}{y} \!\!\! \inty{x_{i}}{x} y^m F_{i,j}(x,y)dxdy \nonumber \\
% &=& \sum_{l=1}^{3} \sum_{k=1}^{3} A^m_{l}(y,y_{j})C_{lk;i,j} \left(x-x_{i}\right)^{k} \left(y-y_{j}\right)^{l},\;\;\; \left(m=0,1,2\right),\label{eq:37}
\end{eqnarray}
}where $\vect{G}^{m} = (G_x^m,G_y^m)$, $G_x^0=G_y^0=G^0$, and $A_k^m$ has been already defined in Equation (\ref{eq:10}). Note that the order of the interpolation function of $f$ is same as that used in the two-dimensional CIP-CSL2 scheme \citep{2002CoPhC.148..137T}.

To determine the coefficients $C_{lk;i,j}$, we use the variables at the upwind position as constraints,
\begin{eqnarray}
\left\{
\begin{array}{lll}
F_{i,j}(x_{i},y_{j}) &=& f_{i,j},\\
F_{i,j}(x_{iup},y_{j}) &=& f_{iup,j},\\
F_{i,j}(x_{i},y_{jup}) &=& f_{i,jup},\\
F_{i,j}(x_{iup},y_{jup}) &=& f_{iup,jup},\\
% G^0_{i,j}(x_{iup},y_{jup}) &=& \sgn \left(\zeta_{i,j}\right) \sgn \left(\eta_{i,j}\right) M^0_{icell,jcell},\\
% G^m_{x;i,j}(x_{iup},y_{jup}) &=& \sgn \left(\zeta_{i,j}\right) \sgn \left(\eta_{i,j}\right) M^m_{x;icell,jcell},\\
% G^m_{y;i,j}(x_{iup},y_{jup}) &=& \sgn \left(\zeta_{i,j}\right) \sgn \left(\eta_{i,j}\right) M^m_{y;icell,jcell},\;\;\;\left(m=1,2\right),\label{eq:38}
\vect{G}^m_{i,j}(x_{iup},y_{jup}) &=& \sgn \left(\zeta_{i,j}\right) \sgn \left(\eta_{i,j}\right) \vect{M}^m_{icell,jcell},\;\;\;\left(m=0,1,2\right),\label{eq:38}
\end{array}
\right.
\end{eqnarray}
where
\begin{eqnarray}
\left\{
\begin{array}{lll}
iup &=& i + \sgn \left(\zeta_{i,j}\right),\\
icell &=& i + \sgn \left(\zeta_{i,j}\right)/2,\\
jup &=& j + \sgn \left(\eta_{i,j}\right),\\
jcell &=& j + \sgn \left(\eta_{i,j}\right)/2,\label{eq:42}
\end{array}
\right.
\end{eqnarray}
are the position of the upwind grid and cell in the $x$ and $y$ directions, and $(\zeta_{i,j},\eta_{i,j})$ is the distance of the upwind departure position relative to $(x_{i},y_{j})$, determined with a second order accuracy,
\begin{eqnarray}
\zeta_{i,j} &=& - u_{j} \Delta t + v_{i} \left(\pdif{u}{y}\right)_{j} \frac{\Delta t^2}{2},\nonumber \\
\eta_{i,j} &=& -v_{i} \Delta t + u_{j} \left(\pdif{v}{x}\right)_{i} \frac{\Delta t^2}{2}.\nonumber
\end{eqnarray}
(Note again that we have assumed $\partial u / \partial x = \partial v / \partial y = 0$, but not necessarily $\partial u / \partial y = \partial v / \partial x = 0$.)
% \begin{eqnarray}
% \zeta_{i,j} &=& -\Delta t \left[u_{i,j} - \frac{\Delta t}{2}\left\{u_{i,j}\left(\pdif{u}{x}\right)_{i,j} + v_{i,j}\left(\pdif{u}{y}\right)_{i,j}\right\}\right],\label{eq:43}\\
% \eta_{i,j} &=& -\Delta t \left[v_{i,j} - \frac{\Delta t}{2}\left\{u_{i,j}\left(\pdif{v}{x}\right)_{i,j} + v_{i,j}\left(\pdif{v}{y}\right)_{i,j}\right\}\right].\label{eq:44}
% \end{eqnarray}
Consequently, the coefficients are explicitly determined, which are listed in \ref{sec:coefficients-mma2d}.

Then, let us consider the time integration of the variables.
For the Vlasov equation in magnetized plasmas, we consider two problems of the advection with constant velocity and the solid body rotation.
\subsection{Time integration for the advection}\label{sec:mma2d-line-advect}
 When the velocity is constant in space, it is easy to implement the semi-Lagrangian method to the two-dimensional scheme, because the volume element of the fluid does not change its shape. The advection phase of Equations (\ref{eq:12})-(\ref{eq:31}) is calculated as
{\footnotesize
\begin{eqnarray}
{}^{n+1}f_{i,j} &=& {}^{n}F_{i,j}(x_i+\zeta,y_j+\eta),\label{eq:14}\\
{}^{n+1}M_{icell,jcell}^{0} &=& \sgn\left(\zeta\right)\sgn\left(\eta\right)\times \nonumber \\
&& \left[\inty{y_j+\eta}{y_{jup}}\!\!\!\inty{x_i+\zeta}{x_{iup}}{}^{n}F_{i,j}(x,y)dxdy \right.\nonumber \\
&& \left. + \inty{y_j+\eta}{y_{jup}}\!\!\!\inty{x_{iup}}{x_{iup}+\zeta}{}^{n}F_{iup,j}(x,y)dxdy\right.\nonumber \\
&& \left. + \inty{y_{jup}}{y_{jup}+\eta}\!\!\!\inty{x_{i}+\zeta}{x_{iup}}{}^{n}F_{i,jup}(x,y)dxdy\right. \nonumber \\
&& \left. + \inty{y_{jup}}{y_{jup}+\eta}\!\!\!\inty{x_{iup}}{x_{iup}+\zeta}{}^{n}F_{iup,jup}(x,y)dxdy\right],\label{eq:39}\\
% {}^{*}M_{x;icell,jcell}^{m} &=& \sgn\left(\zeta\right)\sgn\left(\eta\right)\times \nonumber \\
% && \left[\inty{y_j+\eta}{y_{jup}}\!\!\!\inty{x_i+\zeta}{x_{iup}}\left(x\right)^m{}^{n}F_{i,j}(x,y)dxdy \right.\nonumber \\
% && \left. + \inty{y_j+\eta}{y_{jup}}\!\!\!\inty{x_{iup}}{x_{iup}+\zeta}\left(x\right)^m{}^{n}F_{iup,j}(x,y)dxdy\right.\nonumber \\
% && \left. + \inty{y_{jup}}{y_{jup}+\eta}\!\!\!\inty{x_{i}+\zeta}{x_{iup}}\left(x\right)^m{}^{n}F_{i,jup}(x,y)dxdy\right. \nonumber \\
% && \left. + \inty{y_{jup}}{y_{jup}+\eta}\!\!\!\inty{x_{iup}}{x_{iup}+\zeta}\left(x\right)^m{}^{n}F_{iup,jup}(x,y)dxdy\right],\label{eq:40}\\
{}^{*}\vect{M}_{icell,jcell}^{m} &=& \sgn\left(\zeta\right)\sgn\left(\eta\right)\times \nonumber \\
&& \left[\inty{y_j+\eta}{y_{jup}}\!\!\!\inty{x_i+\zeta}{x_{iup}}\vect{x}^m\cdot{}^{n}F_{i,j}(x,y)dxdy \right.\nonumber \\
&& \left. + \inty{y_j+\eta}{y_{jup}}\!\!\!\inty{x_{iup}}{x_{iup}+\zeta}\vect{x}^m\cdot{}^{n}F_{iup,j}(x,y)dxdy\right.\nonumber \\
&& \left. + \inty{y_{jup}}{y_{jup}+\eta}\!\!\!\inty{x_{i}+\zeta}{x_{iup}}\vect{x}^m\cdot{}^{n}F_{i,jup}(x,y)dxdy\right. \nonumber \\
&& \left. + \inty{y_{jup}}{y_{jup}+\eta}\!\!\!\inty{x_{iup}}{x_{iup}+\zeta}\vect{x}^m\cdot{}^{n}F_{iup,jup}(x,y)dxdy\right],\;\;\;\left(m=1,2\right).\label{eq:41}
\end{eqnarray}
}Here, we consider the case of the CFL number $<1$.
Using Equation (\ref{eq:36}), integrations in the right-hand side of Equations (\ref{eq:39}) and (\ref{eq:41}) are exactly calculated as, e.g.,
{\footnotesize
\begin{eqnarray}
&&\inty{y_j+\eta}{y_{jup}}\!\!\!\inty{x_i+\zeta}{x_{iup}}\vect{x}^m \cdot{}^{n}F_{i,j}(x,y)dxdy\nonumber \\
&=& \left[\inty{y_j}{y_{jup}}\!\!\!\inty{x_i}{x_{iup}} - \inty{y_j}{y_{j}+\eta}\!\!\!\inty{x_i}{x_{iup}} - \inty{y_j}{y_{jup}}\!\!\!\inty{x_i}{x_{i}+\zeta} + \inty{y_j}{y_{j}+\eta}\!\!\!\inty{x_i}{x_{i}+\zeta}\right]\vect{x}^m \cdot{}^{n}F_{i,j}(x,y) dxdy\nonumber \\
&=& \frac{{}^{n}\vect{M}_{icell,jcell}^m}{\sgn\left(\zeta\right)\sgn\left(\eta\right)}-{}^{n}\vect{G}_{i,j}^m(x_{iup},y_{j}+\eta)
-{}^{n}\vect{G}_{i,j}^m(x_i+\zeta,y_{jup})+{}^{n}\vect{G}_{i,j}^m(x_i+\zeta,y_j+\eta).\nonumber
\end{eqnarray}
}% Thus the conservation of mass is guaranteed.
 We next advance the non-advection phase. Since $\vect{u}$ is constant, the right-hand sides of Equations (\ref{eq:30}) and (\ref{eq:31}) are equal to $u M_{x}^{m-1}$ and $v M_{y}^{m-1}$, respectively. Therefore, the non-advection phase is calculated as
{\small
\begin{eqnarray}
{}^{n+1}\vect{M}_{i+1/2,j+1/2}^{1} &=& {}^{*}\vect{M}_{i+1/2,j+1/2}^{1} + \vect{u} {}^{n+1}M_{i+1/2,j+1/2}^{0}\Delta t,\label{eq:55}\\
% {}^{n+1}M_{y;i+1/2,j+1/2}^{1} &=& {}^{*}M_{y;i+1/2,j+1/2}^{1} + v {}^{n+1}M_{i+1/2,j+1/2}^{0}\Delta t,\label{eq:56}\\
{}^{n+1}\vect{M}_{i+1/2,j+1/2}^{2} &=& {}^{*}\vect{M}_{i+1/2,j+1/2}^{2} \nonumber \\
&&+ \vect{u} \left({}^{*}\vect{M}_{i+1/2,j+1/2}^{1}+\vect{u} {}^{n+1}M_{i+1/2,j+1/2}^{0} \frac{\Delta t}{2}\right) \Delta t.\label{eq:57}
% {}^{n+1}M_{y;i+1/2,j+1/2}^{2} &=& {}^{*}M_{y;i+1/2,j+1/2}^{2} \nonumber \\
% &&+ v \left({}^{*}M_{y;i+1/2,j+1/2}^{1}+v {}^{n+1}M_{i+1/2,j+1/2}^{0} \frac{\Delta t}{2}\right) \Delta t.\label{eq:58}
\end{eqnarray}
}% Similar to the MMA1D, the MMA2D with the semi-Lagrangian method can be applied to a large CFL number by replacing $(i,j)$ in Equations (\ref{eq:14})-(\ref{eq:41}) by $(i-{\rm Int}(u \Delta t/|\Delta x|),j-{\rm Int}(v \Delta t/|\Delta y|))$.

{ Similar to the one-dimensional case, in the two-dimensional advection problem with constant velocity, the MMA scheme exactly guarantees the conservation of the zeroth to second order central moments in the $x$ and $y$ directions.}
\subsection{Time integration for the solid body rotation}\label{sec:mma2d-solid-rotation-1}
When the velocity varies in space, it is quite complicated to implement the semi-Lagrangian method, because the exact integration of the interpolation function over a distorted volume element is required. In such a general case (including the solid body rotation problem), it is rather convenient to use the finite volume method for the time integration of the moments. \cite{2007JCoPh.222..849I} implemented a convenient time integration method to the CIP-CSL2 scheme, in which cell-integrated values are advanced by the finite volume method with the Runge-Kutta time integration, whereas point values are advanced by the semi-Lagrangian method. 
% This method is a combination of the CIP-CSL2 and IDO-CF schemes; the former employs the semi-Lagrangian method while the later is the Eulerian method with the Runge-Kutta time integration for the advance of all variables. 

The MMA2D employs their strategy. When the velocity is given as $(u,v) = (-\omega y, \omega x)$, Equations (\ref{eq:29})-(\ref{eq:31}) are written in the following finite volume formulation,
{\scriptsize
\begin{eqnarray}
\frac{\partial M^0_{i+1/2,j+1/2}}{\omega \partial t} &=& -\left\{\left(\int yfdy\right)_{i,j+1/2} - \left(\int yfdy\right)_{i+1,j+1/2} \right\}\nonumber \\
&&+ \left\{\left(\int xfdx \right)_{i+1/2,j} - \left(\int xfdx\right)_{i+1/2,j+1}\right\},\label{eq:67} \\
\frac{\partial M^m_{x;i+1/2,j+1/2}}{\omega \partial t} &=& -\frac{1}{m!} \left\{x^m_{i} \left(\int yfdy\right)_{i,j+1/2} - x^m_{i+1} \left(\int yfdy\right)_{i+1,j+1/2} \right\}\nonumber \\
&&+ \frac{1}{m!} \left\{\left(\int x^{m+1} fdx \right)_{i+1/2,j} - \left(\int x^{m+1} fdx\right)_{i+1/2,j+1}\right\}\nonumber \\
&&- \frac{1}{\left(m-1\right)!} \left(\inty{}{} \!\!\! \inty{}{} y x^{m-1} fdxdy\right)_{i+1/2,j+1/2},\;\;\;\left(m=1,2\right),\label{eq:68}\\
\frac{\partial M^m_{y;i+1/2,j+1/2}}{\omega \partial t} &=& -\frac{1}{m!} \left\{\left(\int y^{m+1} fdy\right)_{i,j+1/2} - \left(\int y^{m+1} fdy\right)_{i+1,j+1/2} \right\}\nonumber \\
&& +\frac{1}{m!} \left\{y^m_{j}\left(\int x fdx \right)_{i+1/2,j} - y^m_{j+1} \left(\int x fdx\right)_{i+1/2,j+1}\right\}\nonumber \\
&& +\frac{1}{\left(m-1\right)!} \left(\inty{}{} \!\!\! \inty{}{} x y^{m-1} fdxdy\right)_{i+1/2,j+1/2},\;\;\;\left(m=1,2\right).\label{eq:69}
\end{eqnarray}
}Evidently, Equation (\ref{eq:67}) guarantees the conservation of mass. Line-integrated variables of $f$ (up to third order) appearing in the right-hand side of Equations (\ref{eq:67})-(\ref{eq:69}) are constructed from the interpolation function as 
% The two dimensional CIP-CSL2 and IDO-CF schemes treat them as dependent variables, while the MMA2D is not. 
% We construct them from the interpolation function,
{\footnotesize
\begin{eqnarray}
\left(\frac{1}{m!} \int x^m fdx\right)_{icell,j} &=& \sgn\left(\zeta_{i,j}\right) \frac{1}{m!} \inty{x_i}{x_{iup}}x^m F_{i,j}(x,y_{j})dx \nonumber \\
 &=& \sgn\left(\zeta_{i,j}\right) \sum_{k=1}^{3} A^m_{k}(x_{iup},x_{i}) C_{1k;i,j} \Delta x^k,\;\;\; \left(m=0,1,2,3\right),\label{eq:62}\\
\left(\frac{1}{m!} \int y^m fdy\right)_{i,jcell} &=& \sgn\left(\eta_{i,j}\right) \frac{1}{m!} \inty{y_j}{y_{jup}}y^m F_{i,j}(x_{i},y)dy \nonumber \\
&=& \sgn\left(\eta_{i,j}\right) \sum_{l=1}^{3} A^m_{l}(y_{jup},y_{j}) C_{l1;i,j} \Delta y^l,\;\;\; \left(m=0,1,2,3\right),\label{eq:63}
\end{eqnarray}
}where we expand the definition of $A_k^m$ for the third order moment,
{\normalsize
\begin{eqnarray}
A_k^3(x,x_i) &=& \frac{1}{\left(k+1\right)\left(k+2\right)\left(k+3\right)} \times \nonumber \\
&& \left\{\frac{k\left(k+1\right)\left(k+2\right)}{6}x^3 + \frac{k\left(k+1\right)}{2}x_i x^2 + k x_i^2 x + x_i^3\right\}.\label{eq:70}
\end{eqnarray}
}The non-advection terms (third terms on the right-hand side) of Equations (\ref{eq:68}) and (\ref{eq:69}) are equal to $-M_{y}^{1}$ and $M_{x}^{1}$ for $m=1$. For $m=2$, the cross moment appears, which is also constructed as
{\footnotesize
\begin{eqnarray}
\left(\inty{}{} \!\!\! \inty{}{} x y fdxdy\right)_{icell,jcell} &=& \sgn\left(\zeta_{i,j}\right) \sgn\left(\eta_{i,j}\right) \inty{y_{j}}{y_{jup}}\!\!\!\inty{x_{i}}{x_{iup}}xy F_{i,j}(x,y)dxdy \nonumber \\
&=& \sgn\left(\zeta_{i,j}\right) \sgn\left(\eta_{i,j}\right) \times \nonumber \\
&& \sum_{l=1}^{3}\sum_{k=1}^{3} A^1_{l}(y_{jup},y_{j})A^1_{k}(x_{iup},x_{i}) C_{lk;i,j} \Delta x^k \Delta y^l.\label{eq:71}
\end{eqnarray}
}

For stable calculation, the time integration of Equations (\ref{eq:67})-(\ref{eq:69}) is implemented with the third-order TVD Runge-Kutta method \citep{1988JCoPh..77..439S,1998MaCom..67...73G}. 
To construct the variables (Equations (\ref{eq:62}), (\ref{eq:63}), and (\ref{eq:71})) at each Runge-Kutta stage, we should calculate the coefficients of the interpolation function at each stage, which depend on $f$ as well as $\vect{M}^{m}$. 
Therefore, intermediate values of $f$ at each stage are necessary. To obtain these, we first solve the equations of the characteristics with the Runge-Kutta method,
\begin{eqnarray}
\dif{\zeta_{i,j}}{t} &=& -\omega \left(y_{j}+\eta_{i,j}\right), \;\;\; {\rm with \;\;} {}^{0}\zeta_{i,j} = 0,\nonumber\\
\dif{\eta_{i,j}}{t} &=& \omega \left(x_{i}+\zeta_{i,j}\right), \;\;\; {\rm with \;\;} {}^{0}\eta_{i,j} = 0,\nonumber
\end{eqnarray}
to get intermediate values of the upwind departure position at each stage, $(x_{i} + {}^{k}\zeta_{i,j}, y_{j} + {}^{k}\eta_{i,j})$, where the left-superscript $k$ denotes the Runge-Kutta stage. 
The solution is exactly obtained as
\begin{eqnarray}
x_{i} + {}^{k}\zeta_{i,j} &=& \cos\left(-{}^{k} \alpha \omega \Delta t\right)x_{i} -\sin \left(-{}^{k} \alpha \omega \Delta t\right) y_{j},\nonumber\\
y_{j} + {}^{k}\eta_{i,j} &=& \sin \left(-{}^{k} \alpha \omega \Delta t\right) x_{i} + \cos\left(-{}^{k} \alpha \omega \Delta t\right) y_{j}\nonumber,
\end{eqnarray}
where ${}^{k} \alpha$ is the Runge-Kutta coefficient, $({}^{1} \alpha,{}^{2} \alpha)=(1,0.5)$ for the third-order TVD method.
Thus, the intermediate values of $f$ are calculated with the semi-Lagrangian method,
\begin{eqnarray}
{}^{k}f_{i,j} &=& {}^n F_{i,j}(x_{i}+{}^k\zeta_{i,j}, y_{j}+{}^k\eta_{i,j}).\nonumber
\end{eqnarray}
Consequently, we can calculate the coefficients at each stage, and then advance the moments with the third-order TVD Runge-Kutta method, which is implemented as
follows,
\begin{eqnarray}
{}^{1}\vect{M}^{m} &=& {}^{n}\vect{M}^{m} + \vect{R}({}^{n}f,{}^{n}\vect{M}^{m})\omega \Delta t,\nonumber \\
{}^{2}\vect{M}^{m} &=& \frac{3}{4} {}^{n}\vect{M}^{m} + \frac{1}{4} \left\{ {}^{1}\vect{M}^{m} + \vect{R}({}^{1}f,{}^{1}\vect{M}^{m})\omega \Delta t\right\},\nonumber \\
{}^{n+1}\vect{M}^{m} &=& \frac{1}{3} {}^{n}\vect{M}^{m} + \frac{2}{3} \left\{ {}^{2}\vect{M}^{m} + \vect{R}({}^{2}f,{}^{2}\vect{M}^{m})\omega \Delta t\right\},\;\;\;(m=0,1,2),\label{eq:16}
\end{eqnarray}
where $\vect{R}$ stands for the right-hand side of Equations (\ref{eq:67})-(\ref{eq:69}).
% Finally with $k=k_{\rm max}$ (the Runge-Kutta order), we advance ${}^{n+1}f_{i,j} = {}^{k_{\rm max}}f_{i,j}$.
Finally, we calculate the point value as
{\small
\begin{eqnarray}
{}^{n+1}f_{i,j} = {}^{n}F_{i,j}(\cos\left(-\omega \Delta t\right)x_{i} -\sin \left(-\omega \Delta t\right) y_{j},\sin \left(-\omega \Delta t\right) x_{i} + \cos\left(-\omega \Delta t\right) y_{j}).\label{eq:60}
\end{eqnarray}
}

{ Unlike the advection problem, the first and second order central moments are not necessarily conserved quantities in the solid body rotation problem. On the other hand, the sum of the square of the first order moments (i.e., the radius of rotation),
\begin{eqnarray}
\left(\inty{-\infty}{\infty}\!\inty{-\infty}{\infty} x f dxdy\right)^2 + \left(\inty{-\infty}{\infty}\!\inty{-\infty}{\infty} y f dxdy\right)^2, \label{eq:59}
\end{eqnarray}
and the sum of the second order moments,
\begin{eqnarray}
\frac{1}{2} \inty{-\infty}{\infty}\!\inty{-\infty}{\infty} \left(x^2+y^2\right) f dxdy,\label{eq:61}
\end{eqnarray}
should be conserved. Equations (\ref{eq:68}) and (\ref{eq:69}) $(m=2)$ guarantee the conservation of Equation (\ref{eq:61}). Taking the sum of Equations (\ref{eq:68}) and (\ref{eq:69}) $(m=1)$ over $(i,j)$ and then implementing the Runge-Kutta calculation, we obtain
\begin{eqnarray}
\sum_{i,j} {}^{n+1} M^{1}_{x} &=& \sum_{i,j}\left\{ \left[1-\frac{\left(\omega \Delta t\right)^2}{2}\right] {}^{n} M^{1}_{x} - \omega \Delta t \left[1-\frac{\left(\omega \Delta t\right)^2}{6}\right] {}^{n} M^{1}_{y} \right\}, \nonumber \\
\sum_{i,j} {}^{n+1} M^{1}_{y} &=& \sum_{i,j}\left\{ \left[1-\frac{\left(\omega \Delta t\right)^2}{2}\right] {}^{n} M^{1}_{y} + \omega \Delta t \left[1-\frac{\left(\omega \Delta t\right)^2}{6}\right] {}^{n} M^{1}_{x} \right\}. \nonumber 
\end{eqnarray}
This is a third-order accurate solution of the centroid position of a profile.
Equation (\ref{eq:59}) is numerically calculated as
\begin{eqnarray}
&& \left( \sum_{i,j} {}^{n+1} M^{1}_{x} \right)^2+ \left( \sum_{i,j} {}^{n+1} M^{1}_{y} \right)^2 \nonumber \\
&& = \left[1-\frac{\left(\omega \Delta t\right)^4}{12}+\frac{\left(\omega \Delta t\right)^6}{36}\right] \left[ \left( \sum_{i,j} {}^{n} M^{1}_{x} \right)^2+ \left( \sum_{i,j} {}^{n} M^{1}_{y} \right)^2 \right].
\end{eqnarray}
Therefore, the MMA scheme will preserve the orbit of rotation with high accuracy, although it does not guarantee the exact conservation.
}

\section{Numerical tests of MMA1D}\label{sec:numer-tests-mma1d}
We perform some numerical tests with the MMA1D, and compare them to the CIP-CSL2 and PFC schemes. 
The accuracy of the CIP-CSL2 scheme is almost same as that of the CIP scheme \citep{2001mwr...129..332Y}. 
Comparisons of the CIP and PFC schemes with others are found in, e.g., \cite{2003CoPhC.150..247F}, and \cite{2008EP&S...60..773U}.
Note that it is important for the MMA scheme to accurately give the initial condition of the piecewise moments.
In the following simulations, we determine the initial condition of the moments by a Gauss quadrature method.
\subsection{Advection problem}\label{sec:linear-advection}
\begin{figure}[htbp]
\centering
\includegraphics[clip,angle=0,scale=.6]{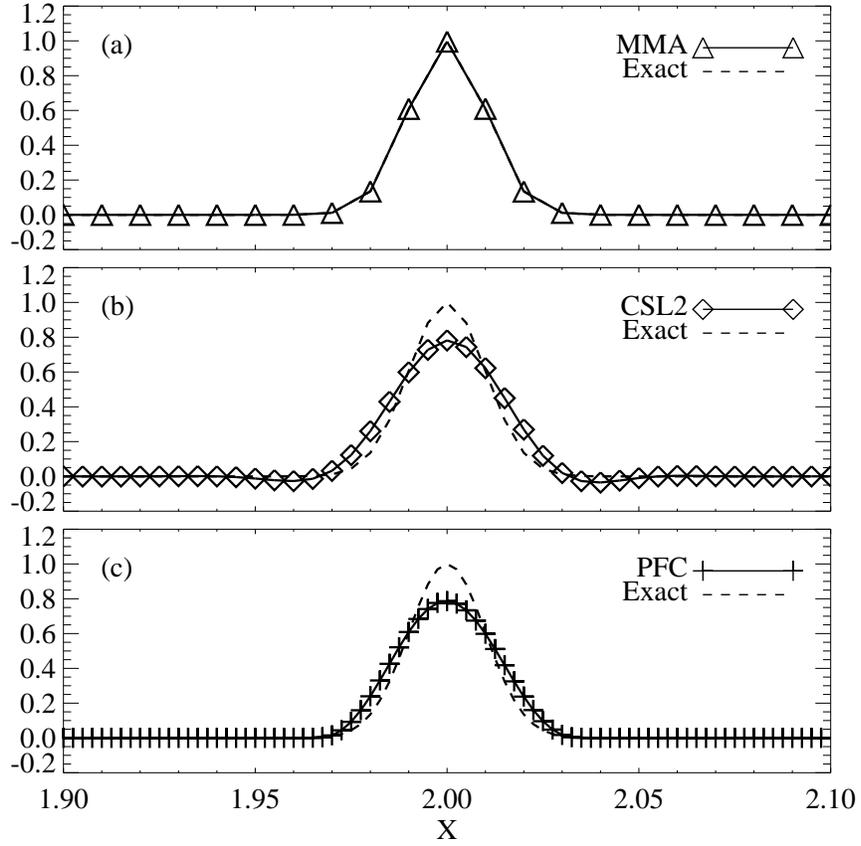}
\caption{Numerical tests of the one-dimensional advection of a gaussian profile with (a) the MMA, (b) the CIP-CSL2, and (c) the PFC schemes. Solid lines with symbols show the simulation result, and dashed lines are the exact solution. In (a), the solid line overlaps the dashed line. The grid sizes are 0.01 for the MMA, 0.005 for the CIP-CSL2, and 0.0025 for the PFC.}
\label{fig:1}
\end{figure}

We test the one-dimensional advection problem with constant velocity,
\begin{eqnarray}
\pdif{f}{t} + u\pdif{f}{x} = 0,
\end{eqnarray}
of a gaussian profile,
\begin{eqnarray}
f(x,t=0) = \exp\left[-\frac{x^2}{2 \sigma_x^2}\right],
\end{eqnarray}
with $u=1$ and $\sigma_x = 0.01$. 
{ Since the numbers of dependent variables are different among the three schemes (four for the MMA, two for the CIP-CSL2, and one for the PFC), we use different grid sizes (0.01 for the MMA, 0.005 for the CIP-CSL2, and 0.0025 for the PFC) so that the total memory usage is equal. The CFL number is 0.2. Figure \ref{fig:1} shows the results at $t=2.0$ solved by the MMA, CIP-CSL2, and PFC schemes. While the CIP-CSL2 and PFC schemes show numerical diffusion, the MMA scheme provides quite an accurate solution.
}
% The grid size and CFL number are 0.005 and 0.2.
% Figure \ref{fig:1} shows the results after 4,000 time steps solved by the MMA, CIP-CSL2, and PFC schemes. The gaussian profile is represented by less than 10 grid points. While the CIP-CSL2 and PFC schemes show considerable numerical diffusion, the MMA scheme provides quite an accurate solution. 
% The local error at the center of the profile, $\epsilon = |f(0,0)-f(ut,t)|$, is $\sim 10^{-4}$ times smaller for the MMA scheme than the others.
% The error of the MMA is $\sim 10^{-4}$ times smaller than that of others.
% The maximum error of the MMA is $\sim 4 \times 10^{-5}$.

\begin{figure}[htbp]
\centering
\includegraphics[clip,angle=0,scale=.6]{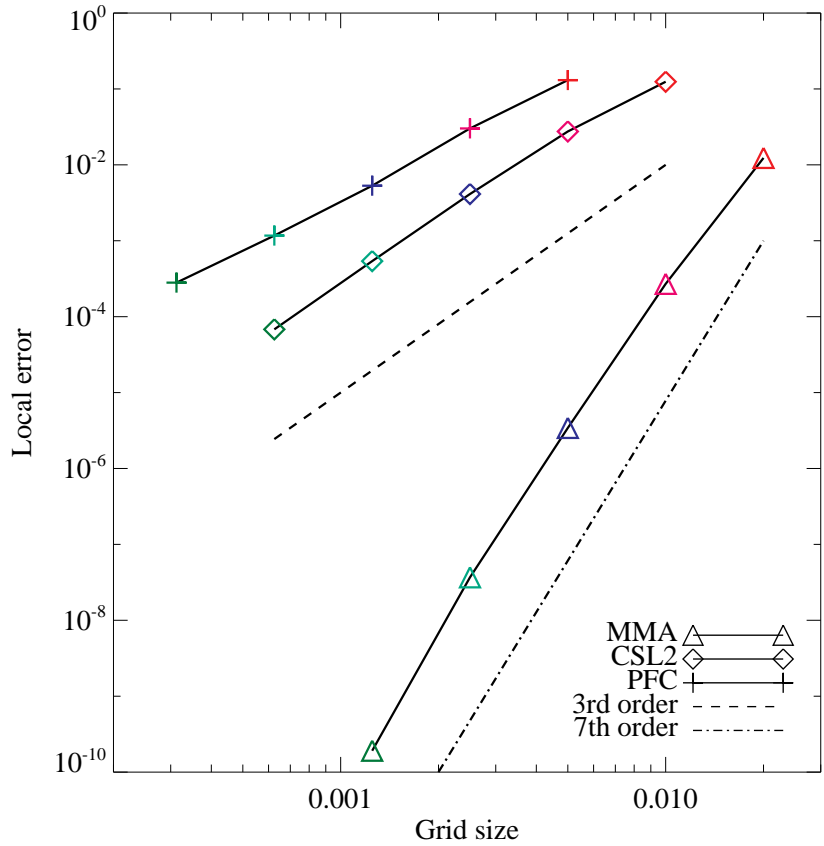}
\caption{Local errors of the one-dimensional advection of a gaussian profile as a function of the grid size. Triangles, diamonds, and crosses are the results obtained from the MMA, CIP-CSL2, and PFC schemes. Dashed and dot-dashed lines represent the third and seventh order accuracies. Symbols with the same colors are obtained from the simulations with the same memory usage.}
\label{fig:2}
\end{figure}

From simulation runs with different grid sizes, we evaluate the order of accuracy of the schemes. Figure \ref{fig:2} shows the local error at the center of the profile, $|f(0,0)-f(ut,t)|$, as a function of the grid size. The slope of a line corresponds to the order of accuracy in space. While the CIP-CSL2 (diamonds) and PFC (crosses) schemes show the third order accuracy (dashed line), the MMA scheme (triangles) achieves the seventh order accuracy (dot-dashed line). 
% Since the MMA scheme treats four dependent variables that are two and four times as much as of the CIP-CSL2 and PFC schemes, the total memory usage is different among the three schemes when the grid size is equal.
In Figure \ref{fig:2}, the simulation results with the same memory usage are denoted as symbols with the same colors.
The error of the MMA scheme is much smaller than the others with the same memory usage.
% Therefore the MMA provides an accurate solution quite effectively. 
% Therefore the MMA provides a solution quite effectively from the viewpoint of the memory usage.
% Considering that the MMA needs four dependent variables that are two and four times as much as of the CIP-CSL2 and PFC, the MMA provides a solution quite effectively from the viewpoint of the memory usage.

% To check the numerical dispersion (phase error) of the MMA, Figure \ref{fig:3} shows the Fourier spectra of the square wave advection. A CFL number 0.1 is used. The wavenumber $k$ and frequency $w$ are normalized by their Nyquist number. 
% All of the three provide a good phase speed. However, the PFC shows considerable dissipation of waves with $k \gsim 0.5$, because this is a relatively diffusive scheme. The CIP-CSL2 seems better than the PFC, but also shows the dissipation  with wavenumber close to the Nyquist number. 
% % Both the MMA (left) and CIP-CSL2 (right) provide a good phase speed. It has been already suggested that in this respect the CIP schemes (including CIP-CSL2) are superior to other schemes such as the Lax-Wendroff, spline, etc. \citep{2001JCoPh.169..556Y}.
% % % However, as can be seen in the right plot, the CIP-CSL2 dissipates the power of waves with wavenumber close to the Nyquist number.
% % However, the dissipation of waves with wavenumbers close to the Nyquist number is seen in the CIP-CSL2.
%  On the other hand, the MMA completely describes waves up to the Nyquist wavenumber with little numerical dispersion nor dissipation.

\begin{figure}[htbp]
\centering
\includegraphics[clip,angle=0,scale=.6]{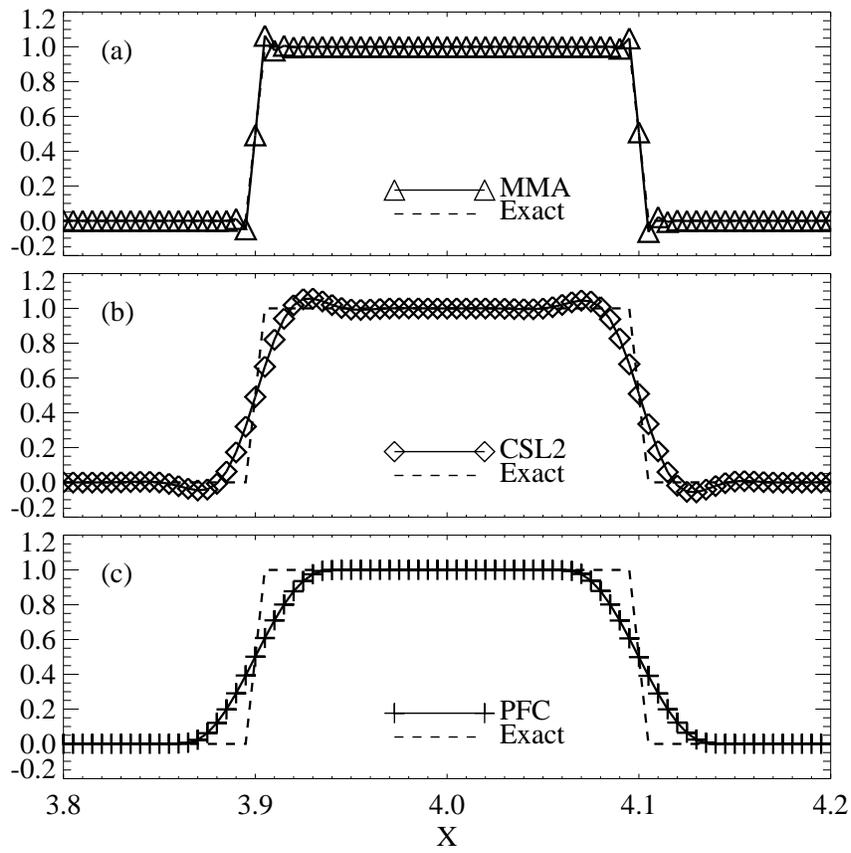}
\caption{Numerical tests of the one-dimensional advection of a square wave profile with (a) the MMA, (b) the CIP-CSL2, and (c) the PFC schemes. Solid lines with symbols show the simulation result, and dashed lines are the exact solution.}
\label{fig:sqr}
\end{figure}

{ Figure \ref{fig:sqr} shows the numerical test of the square wave advection after 4,000 time steps. The grid size and CFL number are 0.005 and 0.2. Compared to the CIP-CSL2 and PFC schemes, the MMA scheme accurately keeps the discontinuity, although it shows small numerical oscillations.
%  Methods to suppress the numerical oscillations will be studied in future.
}
\subsection{Electrostatic Vlasov simulations}\label{sec:es-vlas-simul}
We apply the MMA1D to electrostatic Vlasov-Poisson simulations. The one-dimensional electrostatic Vlasov-Poisson system of equations is written as follows,
\begin{eqnarray}
&& \pdif{f}{t} + v\pdif{f}{x} - \frac{eE}{m} \pdif{f}{v} = 0,\label{eq:75}\\
&& \frac{\partial^2 \phi}{\partial x^2} = 4 \pi n_{0} e \left(\inty{-\infty}{\infty}fdv-1\right),\;\;\;E = -\pdif{\phi}{x},\label{eq:76}
\end{eqnarray}
where $f(v,x)$ is the electron phase space distribution function, $v$ is the electron velocity, $x$ is the position in the configuration space, $e$ is the charge, $m$ is the electron rest mass, $n_{0}$ is the density of ambient immobile ions, and $E(x)$ and $\phi(x)$ are the electrostatic field and potential, respectively.

To solve these equations, we treat eight dependent variables;
%  distribution function, cell-integrated moments in the velocity space, and their cell-integrated values in the configuration space,
 point values of the distribution function, piecewise moments in the velocity space, and their cell-integrated values in the configuration space,
\begin{eqnarray}
f_{i,j}&=&f(v_i,x_j),\nonumber \\
M^{m}_{i+1/2,j} &=& \frac{1}{m!} \inty{v_i}{v_{i+1}} v^m f(v,x_j)dv,\nonumber \\
\tilde{f}_{i,j+1/2}&=&\inty{x_j}{x_{j+1}}f(v_i,x)dx,\nonumber \\
\tilde{M}^{m}_{i+1/2,j+1/2} &=& \frac{1}{m!} \inty{x_j}{x_{j+1}} \!\!\! \inty{v_i}{v_{i+1}} v^m f(v,x)dvdx,\nonumber 
\end{eqnarray}
where the subscripts $i$ and $j$ denote the grid position in the velocity and configuration spaces, and $m=0,1,2$. We then split the Vlasov equation (\ref{eq:75}) into two phases, the advections in the velocity and configuration spaces, which are solved by the MMA and CIP-CSL2 schemes as follows:

(1) The advection equation in the velocity space and its integration over the configuration space are
\begin{eqnarray}
&& \pdif{f_{j}}{t} - \frac{eE_{j}}{m} \pdif{f_{j}}{v} = 0,\nonumber \\
&& \pdif{\tilde{f}_{j+1/2}}{t} - \frac{eE_{j+1/2}}{m} \pdif{\tilde{f}_{j+1/2}}{v} = 0,\nonumber
 \end{eqnarray}
where we approximate $\inty{x_{j}}{x_{j+1}}Efdx \simeq E_{j+1/2} \tilde{f}_{j+1/2}$. 
% The former equation is solved by the MMA for $(f,M^{m})$, and the latter for $(\tilde{f},\tilde{M}^{m})$.
These equations are solved by the MMA scheme for $(f,M^{m})$ and $(\tilde{f},\tilde{M}^{m})$, respectively.

(2) The advection equations in the configuration space,
% \begin{eqnarray}
% \pdif{f}{t} + v\pdif{f}{x} = 0,\nonumber 
% \end{eqnarray}
\begin{eqnarray}
\pdif{f}{t} + v\pdif{f}{x} = 0,\;\;\; \pdif{\tilde{f}}{t} + v\pdif{\tilde{f}}{x} = 0,\nonumber 
\end{eqnarray}
are solved by the CIP-CSL2 scheme for $(f,\tilde{f})$.
Multiplying these equations by $v^{m}/m!$ and then integrating over velocity gives the governing equations of the moments,
% For the advection of moments $M^{m}$, we multiply this equation by $v^{m}/m!$ and then integrate over the velocity, yielding
% \begin{eqnarray}
% \pdif{M^m_{i+1/2}}{t} + \left(m+1\right) \pdif{M^{m+1}_{i+1/2}}{x} = 0.\nonumber 
% \end{eqnarray}
\begin{eqnarray}
\pdif{M^m_{i+1/2}}{t} + \left(m+1\right) \pdif{M^{m+1}_{i+1/2}}{x} = 0, \;\;\; \pdif{\tilde{M}^m_{i+1/2}}{t} + \left(m+1\right) \pdif{\tilde{M}^{m+1}_{i+1/2}}{x} = 0.\nonumber 
\end{eqnarray}
Since the ($m$+1)-th order moment is required for the advance of the $m$-th order one, the system is not closed (the closure problem). 
In this study, we approximate the equations as
% \begin{eqnarray}
% \pdif{M^m_{i+1/2}}{t} + v_{i+1/2} \pdif{M^{m}_{i+1/2}}{x} = 0.\nonumber 
% \end{eqnarray}
\begin{eqnarray}
\pdif{M^m_{i+1/2}}{t} + v_{i+1/2} \pdif{M^{m}_{i+1/2}}{x} = 0, \;\;\;\pdif{\tilde{M}^m_{i+1/2}}{t} + v_{i+1/2} \pdif{\tilde{M}^{m}_{i+1/2}}{x} = 0.\nonumber 
\end{eqnarray}
These advection equations are solved by the CIP-CSL2 scheme for $(M^{0},\tilde{M}^{0})$, $(M^{1},\tilde{M}^{1})$, and $(M^{2},\tilde{M}^{2})$. 
The Poisson equation (\ref{eq:76}) is solved by the Fourier transform.
 The electric charge density on the right-hand side can be immediately calculated by taking the sum of $M^0$ or $\tilde{M}^0/\Delta x$ over the velocity space. Since the MMA scheme guarantees the conservation of $M^0$ and $\tilde{M}^0$ (see Equation~(\ref{eq:23})), and the CIP-CSL2 scheme guarantees the conservation of $\tilde{f}$ and $\tilde{M}^{m}$, our simulation exactly conserves $\tilde{M}^0$. Therefore, we use $\tilde{M}^0$ for the charge density calculation. 
% The Poisson equation (\ref{eq:76}) is solved with the standard method such as the spectral or SOR method. The electric charge density in the right-hand side can be immediately obtained by taking the sum of $M^0_{i+1/2,j}$ or $\tilde{M}^0_{i+1/2,j+1/2}/\Delta x$ in the velocity space. Since the MMA guarantees the conservation of $M^0_{i+1/2,j}$ and $\tilde{M}^0_{i+1/2,j}$ (Equation~(\ref{eq:23})), and the CIP-CSL2 guarantees the conservation of $\tilde{f}_{i,j+1/2}$ and $\tilde{M}^{0}_{i+1/2,j+1/2}$, our simulation exactly conserves $\tilde{M}^0_{i+1/2,j+1/2}$. Therefore it is more suitable to use $\tilde{M}^0_{i+1/2,j+1/2}$ than $M^0_{i+1/2,j}$ for the charge density calculation. 

Physical variables in the system are normalized as follows; velocity by the thermal velocity $v_{th}$, time by the inverse electron plasma frequency $\omega_p^{-1} = \sqrt{m/4 \pi n_{0} e^2}$, and position by the Debye length $\lambda_{D} = v_{th}/\omega_p$.
% The boundary conditions are periodic in the configuration space (except in Section \ref{sec:electr-beam-prop}), and open in the velocity space where constant incoming fluxes are assumed while outgoing fluxes are perfectly lost. 
The boundary conditions are periodic in the configuration space, and open in the velocity space where constant incoming fluxes are assumed while outgoing fluxes are perfectly lost. 
We employ the splitting method of \cite{1976JCoPh..22..330C} for the time advance of the system.
% The time advance of the system is implemented as follows \citep{1976JCoPh..22..330C};
%  (1) advection in the configuration space with a half time step $\Delta t/2$, (2) calculating the charge density and solving the Poisson equation to obtain the electric field, (3) advection in the velocity space with a full time step $\Delta t$, (4) advection in the configuration space with a half time step.  

\subsubsection{Linear Landau damping}\label{sec:line-land-damp}
\begin{figure}[htbp]
\centering
\includegraphics[clip,angle=0,scale=.6]{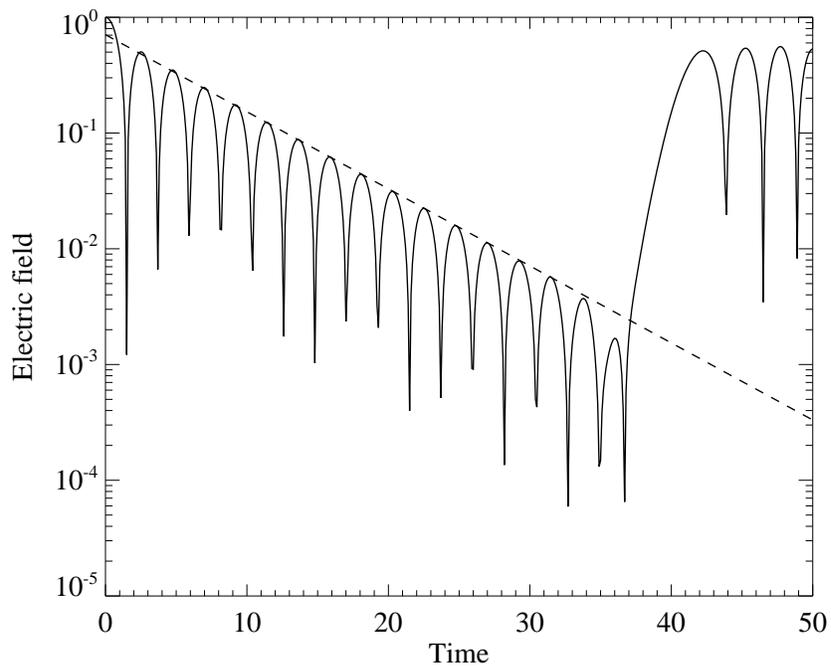}
\caption{Time profile of the electric field (normalized by the initial value) in the linear Landau damping simulation with the MMA scheme. A solid line shows the simulation result, and a dashed line indicates the linear damping rate. The time is normalized by the inverse electron plasma frequency.}
\label{fig:4}
\end{figure}

We first test the linear Landau damping. The initial condition of the electron distribution function is given as follows,
\begin{eqnarray}
f(v,x,t=0) = \frac{1}{\sqrt{2 \pi}} \exp\left(-\frac{v^2}{2}\right) \left[1+0.01\cos\left(0.5x\right)\right].\label{eq:80}
\end{eqnarray}
The simulation domain is $[-4.5,4.5]$ in the velocity space with 32 grid points, and $[0,4\pi]$ in the configuration space with 32 grid points. The time step is 0.01. Figure \ref{fig:4} is the time profile of the electric field (normalized by the initial value), showing the exponential decrease. The simulation (solid line) agrees well with the linear theory (dashed line). The recurrence effect occurs at $t \sim 42$, which is comparable to the theoretical value $t=44.68$ obtained from the free-streaming equation \citep{1976JCoPh..22..330C}.

\subsubsection{Two stream instability}\label{sec:two-stre-inst}
We next test the two stream instability. The initial condition of the electron distribution function is given as follows,
\begin{eqnarray}
f(v,x,t=0) = \frac{1}{\sqrt{2 \pi}} v^2 \exp\left(-\frac{v^2}{2}\right) \left[1+0.05\cos\left(0.5x\right)\right].\label{eq:81}
\end{eqnarray}
The simulation domain is $[-5.0,5.0]$ in the velocity space with 32, 64, or 128 grid points, and $[0,4\pi]$ in the configuration space with 64 grid points. The time step is 0.01. For comparison, we also perform the simulation in which the advections in the velocity space as well as the configuration space are solved by the CIP-CSL2 scheme.

\begin{figure}[htbp]
\centering
\includegraphics[clip,angle=0,scale=.55]{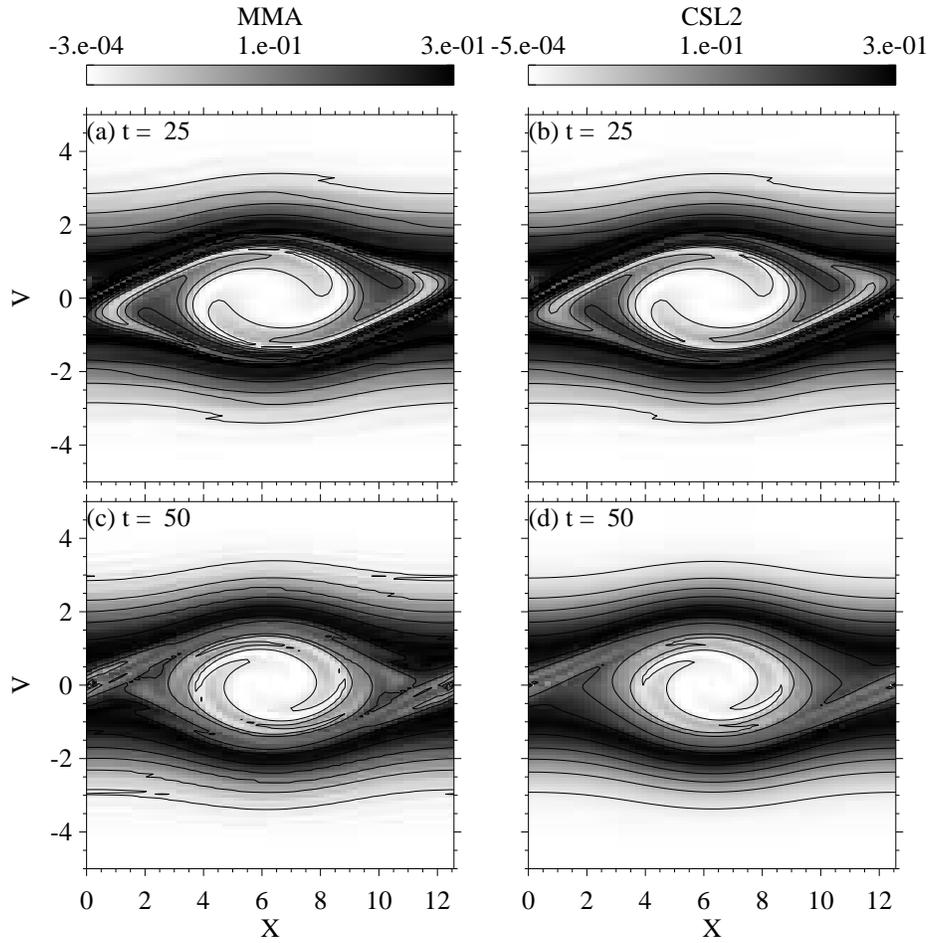}
\caption{Electron phase space distributions (negative images with contours) at (a,b) $t=25$ and (c,d) $t=50$ in the two stream instability simulation, with (a,c) the MMA and (b,d) the CIP-CSL2 schemes. The numbers of grid points are 64 and 128 in the $x$ and $v$ directions. The time, position, and velocity are normalized by the inverse electron plasma frequency, Debye length, and thermal velocity, respectively.}
\label{fig:5}
\end{figure}

\begin{figure}[htbp]
\centering
\includegraphics[clip,angle=0,scale=.55]{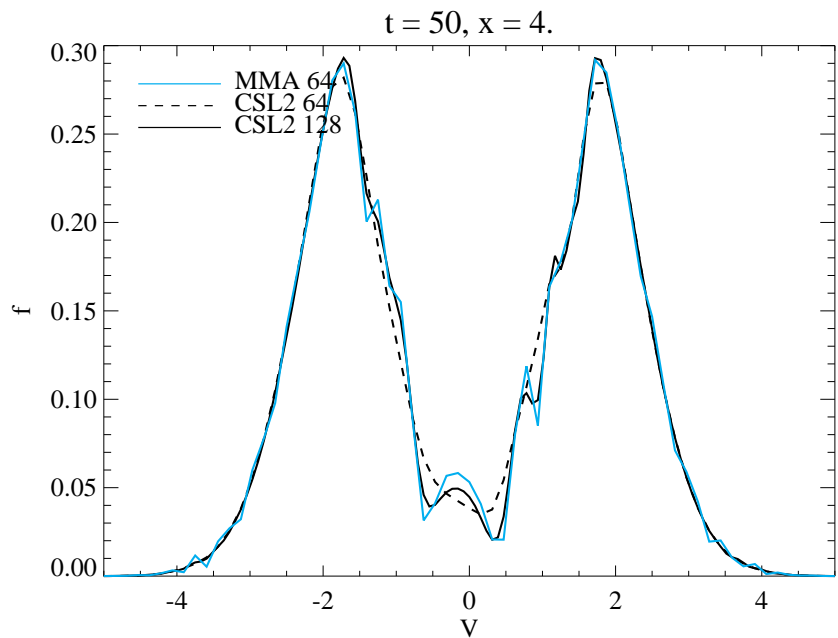}
\caption{Electron velocity space distribution at $x=4.0$ and $t=50$ in the two stream instability simulation. A blue line is obtained from the MMA scheme with 64 grid points in the velocity space. Black solid and dashed lines are from the CIP-CSL2 scheme with 128 and 64 points, respectively. The time, position, and velocity are normalized by the inverse electron plasma frequency, Debye length, and thermal velocity, respectively.}
\label{fig:1dcut}
\end{figure}

Figure \ref{fig:5} shows the electron distribution function at $t=25$ and $t=50$ simulated with the MMA and CIP-CSL2 schemes. Both simulations show the well-known electron hole structure in the phase space. Figure \ref{fig:1dcut} shows the velocity space distribution taken at $x=4.0$ and $t=50$. The MMA scheme (blue line) describes finer structures than the CIP-CSL2 scheme with the same grid points (dashed line), and shows a similar result to the CIP-CSL2 scheme with the double number of grid points (solid line). 
% The MMA describes finer structures than the CIP-CSL2. 
% We confirm that the amplitude of the negative value is smaller for the MMA, despite that the MMA shows more finite structure. 

\begin{figure}[htbp]
\centering
\includegraphics[clip,angle=0,scale=.5]{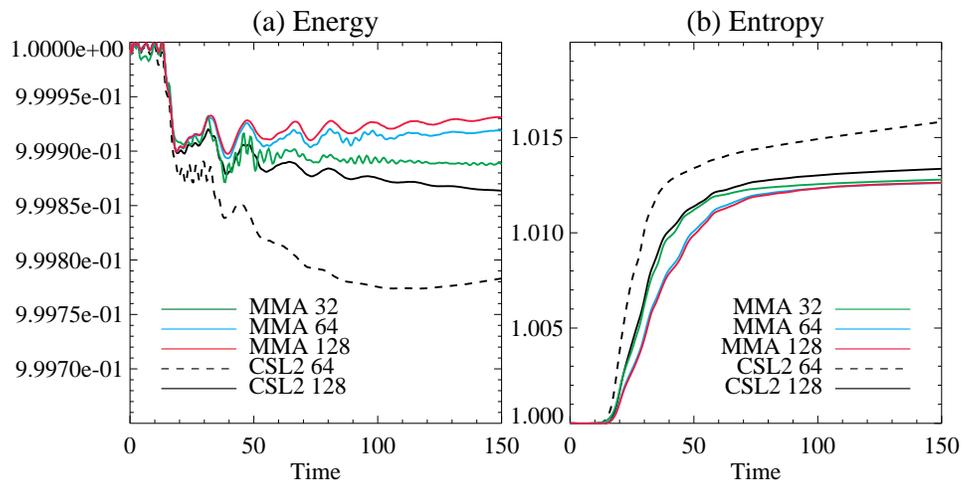}
\caption{Temporal variation of (a) the total energy and (b) the entropy (normalized by their initial values) in the two stream instability simulation. Color lines are obtained from the MMA scheme with the different numbers of grid points in the velocity space (green = 32, blue = 64, and red = 128), and black lines are from the CIP-CSL2 scheme (dashed = 64 and solid = 128 points in the velocity space). The time is normalized by the inverse electron plasma frequency.}
\label{fig:6}
\end{figure}

\begin{figure}[htbp]
\centering
\includegraphics[clip,angle=0,scale=.6]{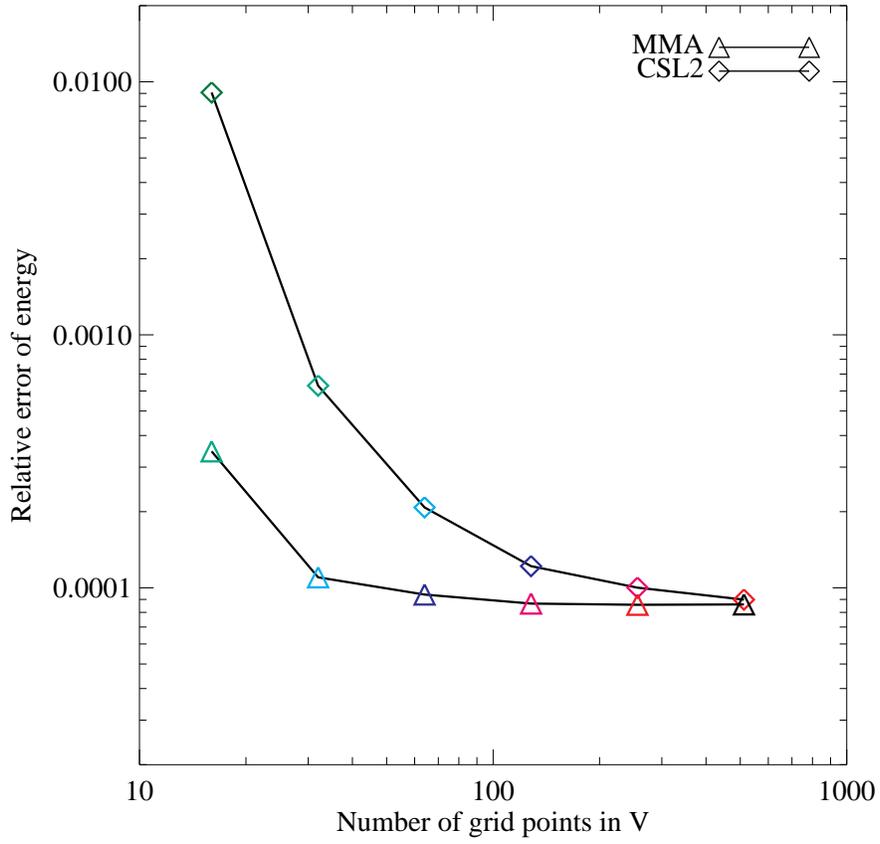}
\caption{Relative error of the total energy at $t=75$, as a function of the number of grid points in the velocity space. Triangles and diamonds are the results obtained from the MMA and CIP-CSL2 schemes. Symbols with the same colors are obtained from the simulations with the same memory usage.}
\label{fig:twst_err}
\end{figure}

In Figure \ref{fig:6}, we check the conservation of the total energy and the entropy,
\begin{eqnarray}
\varepsilon &=& \inty{}{}\left(\frac{E^2}{8\pi}+\inty{}{}\frac{v^2}{2}fdv\right)dx,\nonumber \\
H &=& \inty{}{}\!\!\!\inty{}{}f\left(1-f\right)dvdx,\nonumber
\end{eqnarray}
 with the different schemes (MMA or CIP-CSL2) and the different numbers of grid points (32, 64, or 128) in the velocity space.
%  All simulations show very good results. 
The MMA scheme provides the energy and entropy conservations well, even with the smallest number of grid points (32, green lines). It is noteworthy that the MMA scheme with 32 grid points shows a better result than the CIP-CSL2 scheme with 128 grid points (solid lines), which uses the memory twice as much as the MMA. 
% In this regard, the MMA is an effective scheme.
{ Figure \ref{fig:twst_err} shows the relative error of the total energy $(\varepsilon(t)-\varepsilon(0))/\varepsilon(0)$ at $t=75$ obtained from the both schemes, as a function of the number of grid points in the velocity space. With the sufficiently large number of grid points, the both schemes show the same overall accuracy, because the same method is applied to solve the advection in the configuration space.
% The MMA scheme shows that the error is very close to the minimum value even with the small number of grid points (e.g., light blue symbol).
}

\section{Numerical tests of MMA2D}\label{sec:numer-tests-mma2d}
We perform some numerical tests with the MMA2D, and compare them to the CIP-CSL2 and PFC schemes. We also apply the MMA2D to electromagnetic Vlasov-Maxwell simulations and compare them to the PIC method.
%  We show that the MMA has great advantages of the memory usage as well as the accuracy over other schemes especially for multidimensions, and is quite a powerful scheme for the electromagnetic Vlasov simulation.

\subsection{Solid body rotation problem}\label{sec:solid-rotation}
We test long time simulations of the two-dimensional solid body rotation problem,
\begin{eqnarray}
\frac{\partial f}{\partial t}- y\pdif{f}{x}+ x\pdif{f}{y}=0,\label{eq:79}
\end{eqnarray}
of a gaussian profile,
\begin{eqnarray}
f\left(x,y,t=0\right) = \exp\left[-\frac{\left(x-x_0\right)^2}{2 \sigma_x^2}-\frac{\left(y-y_0\right)^2}{2 \sigma_y^2}\right].\label{eq:77}
\end{eqnarray}
% This problem is crucially important for the electromagnetic Vlasov simulation (gyro motion in the velocity space) and hence the most important target of the MMA. 
% We perform the simulations with the MMA2D, two dimensional CIP-CSL2 \citep{2002CoPhC.148..137T}, and backsubstitution \citep{2006CoPhC.175...86S} schemes.
We compare the simulations with the MMA to the two-dimensional CIP-CSL2 \citep{2002CoPhC.148..137T} and backsubstitution \citep{2006CoPhC.175...86S} schemes.
Since the numbers of dependent variables are different among these three schemes (six for the MMA, four for the CIP-CSL2, and one for the backsubstitution), we use the different numbers of grid points ($34\times34$ for the MMA, $42\times42$ for the CIP-CSL2, and $84\times84$ for the backsubstitution) so that the total memory usage is equal.
The simulation domain is [-0.5,0.5] in both the $x$ and $y$ directions. The open boundary condition is employed where constant incoming fluxes are assumed while outgoing fluxes are perfectly lost. The time steps are 0.004$\pi$ for the MMA and CIP-CSL2 schemes, and 0.002$\pi$ for the backsubstitution scheme so that the CFL number is close among the three simulations.

\begin{figure}[htbp]
\centering
\includegraphics[clip,angle=0,scale=.55]{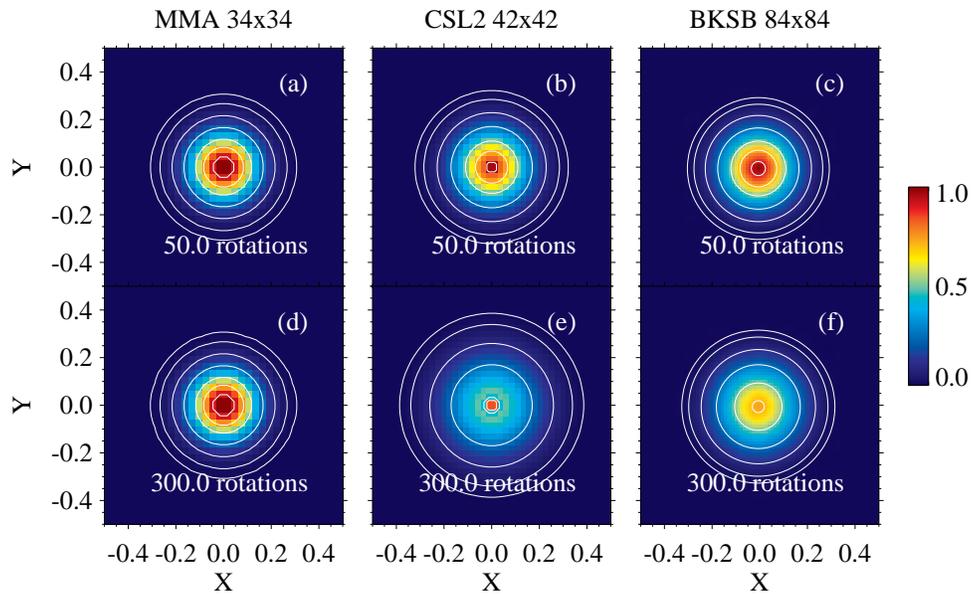}
\caption{Two-dimensional solid body rotation problem of a symmetric gaussian profile after (a,b,c) 50 and (d,e,f) 300 rotations, calculated with (a,d) the MMA, (b,e) the CIP-CSL2, and (c,f) the backsubstitution schemes. Contour levels are [0.01,0.03,0.1,0.25,0.5,0.7,0.9]. The numbers of grid points are $34\times34$ for the MMA, $42\times42$ for the CIP-CSL2, and $84\times84$ for the backsubstitution.}
\label{fig:7}
\end{figure}

Figure \ref{fig:7} shows the results for a symmetric gaussian profile with $x_0=y_0=0,\sigma_x=\sigma_y=0.1$ after 50 and 300 rotations.
% The top and bottom images show the profiles after 50 and 300 rotations, and the left, center, and right images are simulated with the MMA, CIP-CSL2, and backsubstitution, respectively. 
% Note that since the MMA treats six dependent variables while the CIP-CSL2 treats four, we set the different number of grids between the two simulations (CIP-CSL2 uses many grids) to uniform the total memory usage. 
% While the CIP-CSL2 and backsubstitution shows considerable numerical diffusion, the MMA provides quite an accurate solution without the diffusion even after hundreds rotations.
The CIP-CSL2 scheme shows the most serious numerical diffusion. 
The backsubstitution scheme is better than the CIP-CSL2, but also shows considerable diffusion after about a hundred of rotations. 
On the other hand, the MMA scheme provides quite an accurate solution with little diffusion even after hundreds of rotations.

% In this test, the CPU time of the MMA is about three times longer than others, because the MMA constructs the interpolation function at each Runge-Kutta stage (see Section \ref{sec:mma2d-solid-rotation-1}). Then we test the same problem for the CIP-CSL2 and PFC with one-third time steps, and confirm that the results are almost the same as in Figure \ref{fig:7}. Therefore the MMA is quite an effective scheme also in multidimensions.

\begin{figure}[htbp]
\centering
\includegraphics[clip,angle=0,scale=.55]{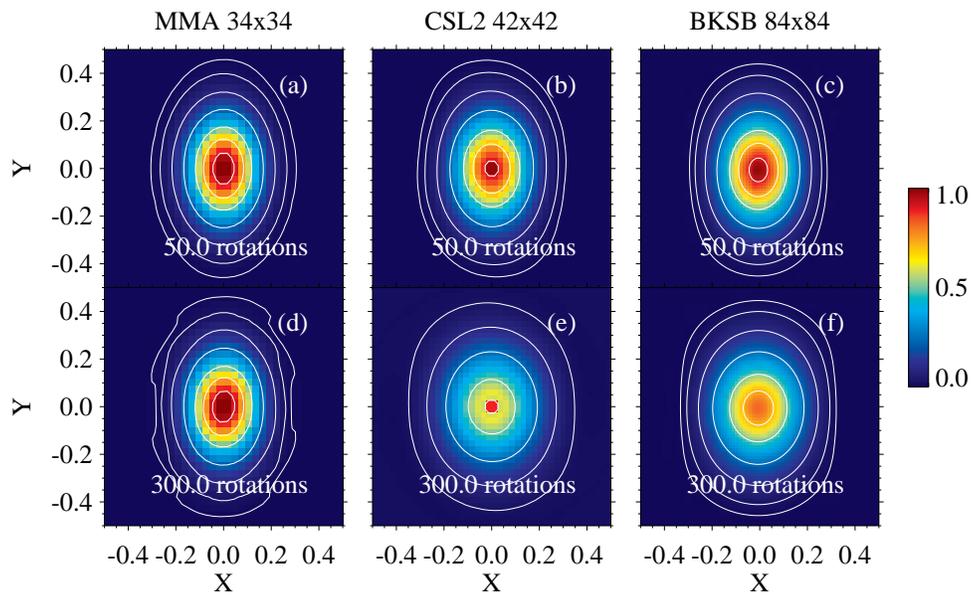}
\caption{Two-dimensional solid body rotation problem of an asymmetric gaussian profile. The format is same as Figure \ref{fig:7}.}
\label{fig:8}
\end{figure}

Figure \ref{fig:8} shows the results for an asymmetric gaussian profile with $x_0=y_0=0,\sigma_x=0.1,\sigma_y=0.15$. The MMA scheme provides a better solution than the others. The gaussian centroid $(x_0,y_0)$ and the standard deviation $(\sigma_x,\sigma_y)$ calculated by the MMA scheme are kept correct even after hundreds of rotations, although a small error is seen on the tail of the profile (the outermost contour). It is also found from this simulation that the phase error of the rotation is very small.

We next test the solid body rotation with the advection problem,
\begin{eqnarray}
\frac{\partial f}{\partial t}-\left(y-y_0\right)\pdif{f}{x}+\left(x-x_0\right)\pdif{f}{y}=0,\label{eq:82}
\end{eqnarray}
which describes the rotation around $(x,y)=(x_0,y_0)$. For magnetized plasmas, this corresponds to the electric field $(\vect{E}\times\vect{B})$ drift motion. To solve this equation, we split it into two phases, the rotation phase (Equation (\ref{eq:79})), and the advection phase,
\begin{eqnarray}
\frac{\partial f}{\partial t}+y_0\pdif{f}{x}-x_0\pdif{f}{y}=0.\nonumber
\end{eqnarray}

\begin{figure}[htbp]
\centering
\includegraphics[clip,angle=0,scale=.55]{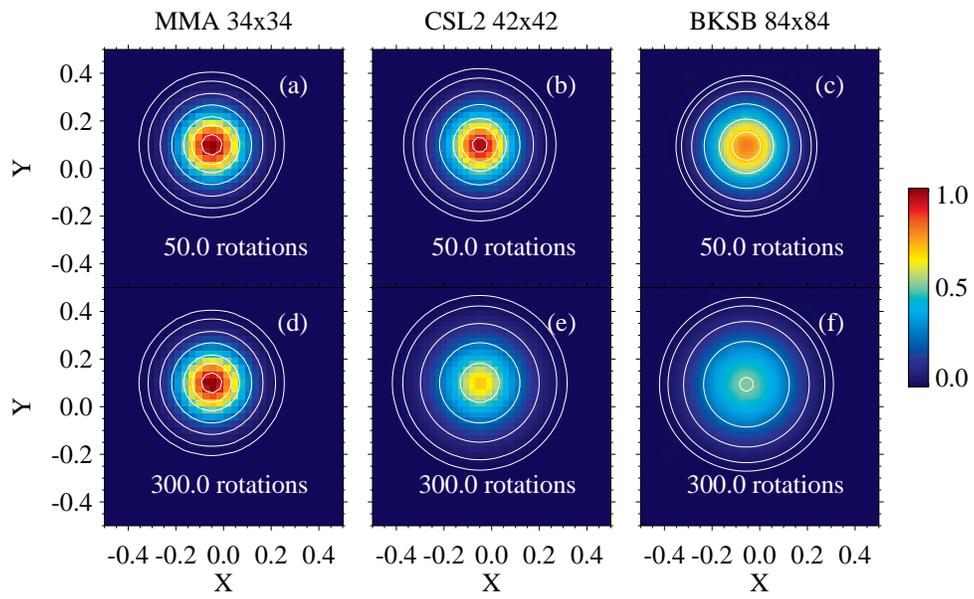}
\caption{Two-dimensional solid body rotation and advection problem of a symmetric gaussian profile. The format is same as Figure \ref{fig:7}.} 
\label{fig:9}
\end{figure}

\begin{figure}[htbp]
\centering
\includegraphics[clip,angle=0,scale=.6]{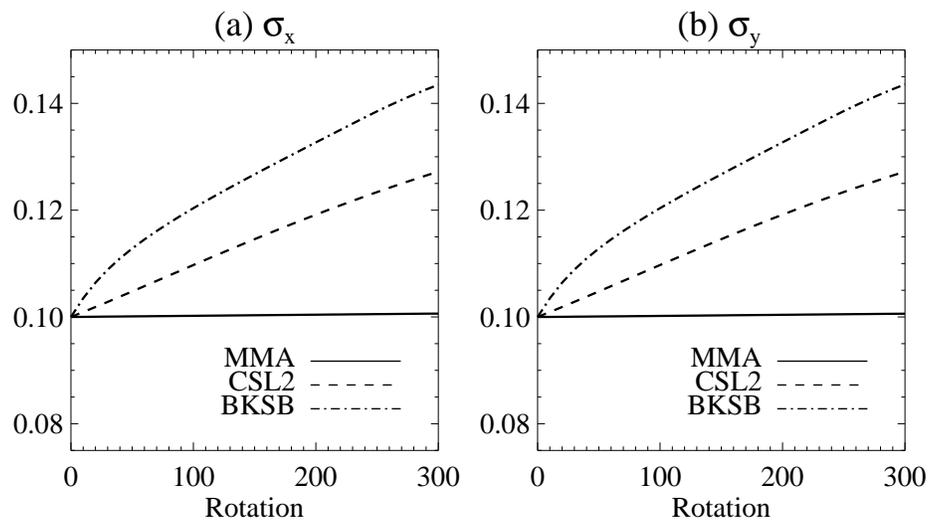}
\caption{Temporal variation of the standard deviations (a) $\sigma_x$ and (b) $\sigma_y$ in the two-dimensional rotation and advection problem shown in Figure \ref{fig:9}. Solid, dash, and dot-dashed lines are obtained from the MMA, CIP-CSL2, and backsubstitution schemes.}
\label{fig:a}
\end{figure}

Figure \ref{fig:9} shows the results for a symmetric gaussian profile with $x_0=-0.05,y_0=0.1,\sigma_x=\sigma_y=0.1$. 
In the problem of the rotation with advection, the backsubstitution scheme shows the most serious diffusion after several tens of rotations.
Again, the MMA scheme provides a better solution with keeping $(x_0,y_0)$ and $(\sigma_x,\sigma_y)$ constant. Figure \ref{fig:a} shows the temporal variation of $(\sigma_x,\sigma_y)$ obtained by fitting the profile with the gaussian function. This result indicates that the MMA scheme can accurately solve the $\vect{E}\times\vect{B}$ drift motion for a long time with little numerical heating. 

\begin{figure}[htbp]
\centering
\includegraphics[clip,angle=0,scale=.6]{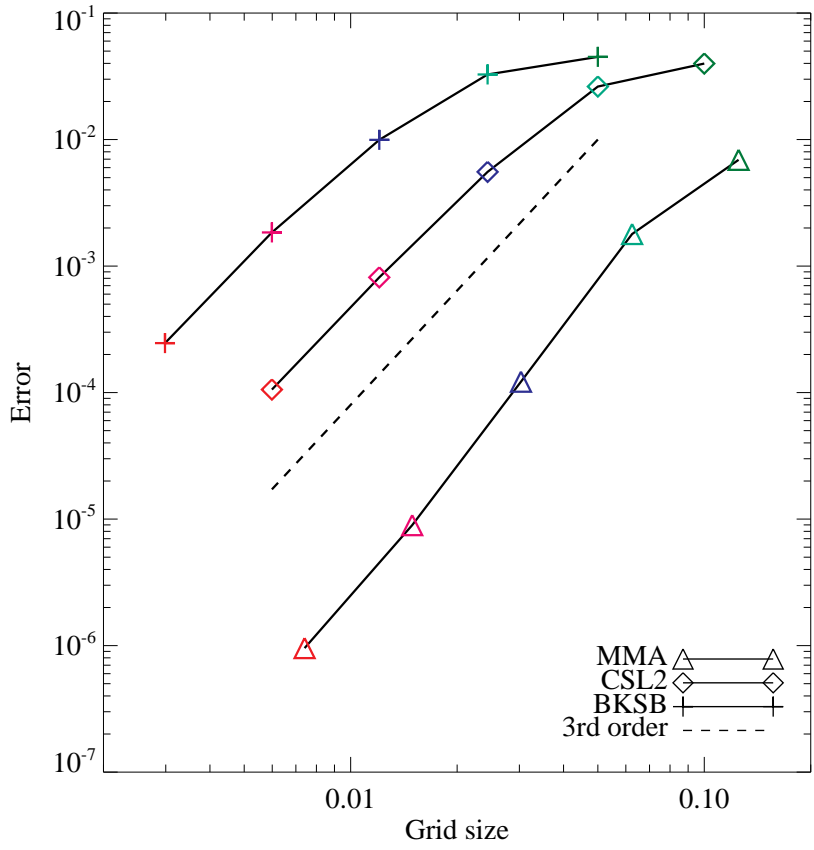}
\caption{Errors of the two-dimensional solid body rotation and advection problem of a symmetric gaussian profile as a function of the grid size. Triangles, diamonds, and crosses are the results obtained from the MMA, CIP-CSL2, and backsubstitution schemes. A dashed line represents the third order accuracy. Symbols with the same colors are obtained from the simulations with the same memory usage.}
\label{fig:err2d}
\end{figure}

{ From these simulation runs with different grid sizes, we evaluate the order of accuracy of the schemes. Figure \ref{fig:err2d} shows the error $\sum_{i,j} |f(x,y,t)-f(x,y,0)|/N$ after 50 rotations, as a function of the grid size ($N$ is the number of grid points). All schemes show the third order accuracy in space (dashed line). In Figure \ref{fig:err2d}, the simulation results with the same memory usage are denoted as symbols with the same colors. The error of the MMA scheme (triangles) is $\sim 10^{-2}$ times smaller than the others with the same memory usage. The errors of the schemes increase with time.
}

We note that the order of the interpolation function of the MMA scheme (Equation (\ref{eq:35})) is exactly same as that of the CIP-CSL2 scheme. Nevertheless, significant differences are found between the two schemes, suggesting that selection of constraints for the interpolation function is important to improve the accuracy. 
In this respect, we consider that the MMA scheme has the reasonable selection on the basis of the concept of the entropy conservation (Section \ref{sec:concept-mma-scheme}), as it is designed specifically for Vlasov simulations.
% In this regard, we consider that our selection based on the conservation of the entropy (Section \ref{sec:concept-mma-scheme}) is much rational. 

\subsection{Electromagnetic Vlasov simulations}\label{sec:em-vlas-simul}
We apply the MMA2D to electromagnetic Vlasov-Maxwell simulations. The one-dimensional electromagnetic Vlasov-Maxwell system of equations is written as follows,
\begin{eqnarray}
&& \pdif{f_s}{t}+v_x\pdif{f_s}{x}+\frac{q_s}{m_s}\left(\vect{E}+\frac{\vect{v}\times\vect{B}}{c}\right)\cdot\pdif{f_s}{\vect{v}} = 0,\;\;\;\left(s=p,e\right),\label{eq:84}\\
&& \pdif{\vect{E}}{t}=c \nabla \times \vect{B} - 4 \pi \vect{j},\label{eq:85}\\
&& \pdif{\vect{B}}{t}=-c \nabla \times \vect{E},\label{eq:86}\\
&& \vect{j} = \sum_{s=p,e}q_s \inty{}{}\vect{v}f_s d\vect{v},\label{eq:87}
\end{eqnarray}
where $\vect{E}$ and $\vect{B}$ are the electric and magnetic fields, $\vect{j}$ is the current density, $c$ is the speed of light, $q_s$ is the charge, and the subscript $s$ denotes particle species ($p$ for protons and $e$ for electrons). We assume the two dimensionality in the velocity space, $\vect{v} = (v_x,v_y,0)$, $\vect{E} = (E_x,E_y,0)$, and $\vect{B} = (0,0,B_z)$.

To solve these equations, we treat twelve dependent variables for both protons and electrons; point values of the distribution function, piecewise moments in the velocity space, and their cell-integrated values in the configuration space,
\begin{eqnarray}
f_{i,j,k}&=&f(v_{x,i},v_{y,j},x_k),\label{eq:88}\nonumber \\
\vect{M}^m_{i+1/2,j+1/2,k} &=& \frac{1}{m!} \inty{v_{y,j}}{v_{y,j+1}} \!\!\! \inty{v_{x,i}}{v_{x,i+1}} \vect{v}^m f(v_x,v_y,x_k)dv_xdv_y,\label{eq:90}\nonumber \\
\tilde{f}_{i,j,k+1/2}&=&\inty{x_k}{x_{k+1}}f(v_{x,i},v_{y,j},x)dx,\label{eq:92}\nonumber \\
\vect{\tilde{M}}^m_{i+1/2,j+1/2,k+1/2} &=& \frac{1}{m!} \inty{x_{k}}{x_{k+1}}\!\!\! \inty{v_{y,j}}{v_{y,j+1}} \!\!\! \inty{v_{x,i}}{v_{x,i+1}} \vect{v}^m f(v_x,v_y,x)dv_xdv_ydx,\label{eq:94}\nonumber
\end{eqnarray}
where the subscripts $i$, $j$, and $k$ denote the grid position in the $v_x$, $v_y$, and $x$ directions, $\vect{M}^{m}=(M^{m}_{v_x},M^{m}_{v_y})$, $M^0_{v_x} = M^0_{v_y}=M^0$, $\vect{\tilde{M}}^{m}=(\tilde{M}^{m}_{v_x},\tilde{M}^{m}_{v_y})$, $\tilde{M}^0_{v_x} = \tilde{M}^0_{v_y}=\tilde{M}^0$, and $m=0,1,2$. 
We then split the Vlasov equation (\ref{eq:84}) into three phases, the advection in the configuration space, the acceleration by electric fields, and the gyro motion \citep{2003JCoPh.186...47G,2009CoPhC.180..365U,2010PhPl...17e2311U}. The acceleration and the gyro motion are solved by the MMA scheme (Sections \ref{sec:mma2d-line-advect} and \ref{sec:mma2d-solid-rotation-1}) for $(f,\vect{M}^m)$ and $(\tilde{f},\vect{\tilde{M}}^m)$. The advection in the configuration space is solved by the CIP-CSL2 scheme for $(f,\tilde{f})$, $(M^0,\tilde{M}^0)$, $(\vect{M}^1,\vect{\tilde{M}}^1)$, and $(\vect{M}^2,\vect{\tilde{M}}^2)$. Those are carried out in the same manner as the electrostatic Vlasov simulation that avoids the closure problem (see Section \ref{sec:es-vlas-simul}). 

% We also split the Maxwell equation into two phases, the advection and non-advection phases. Taking the sum and difference of Equations (\ref{eq:85}) and (\ref{eq:86}) with neglecting the current term, the advection phase is written as simple linear advection equations,
% \begin{eqnarray}
% && \pdif{\left(E_y \pm B_z\right)}{t} \pm c\pdif{\left(E_y \pm B_z\right)}{x} = 0.\label{eq:97}
% \end{eqnarray}
% We employ the CIP-CSL2 to solve them (the application of the CIP to the Maxwell equation is originally proposed by \cite{2006CiCP.1..311O}).  The non-advection phase (current term) is advanced before and after the advection phase with each half time step. 
We solve the Maxwell equation with the implicit scheme \citep{1986HMPhDT...,1987JGR....92.7368H}.
% We solve the Maxwell equation with the implicit scheme \citep{1986HMPhDT...,1987JGR....92.7368H} in Sections \ref{sec:perp-wave-prop} and \ref{sec:harris-current-sheet}, and with the CIP scheme \citep{2006CiCP.1..311O} in Section \ref{sec:perp-shock-waves}.
The current density can be immediately calculated from the sum of $\vect{M^1}$ or $\vect{\tilde{M}}^1/\Delta x$ over the velocity space. We use the latter for the current density calculation, because they are conserved in the configuration space (but not in the velocity space).

Physical variables in the system are normalized as follows; velocity by the speed of light, time by the inverse electron plasma frequency $\omega_{pe}^{-1} = \sqrt{m_e/4 \pi n_e q_e^2}$, electromagnetic fields by an ambient magnetic field, and position by the Debye length $\lambda_{D}$ or the electron inertia length $d_e = c/\omega_{pe}$ depending on problems.
%  The boundary conditions are periodic in the configuration space (except in Section \ref{sec:perp-shock-waves}) and open in the velocity space.
 The boundary conditions are periodic in the configuration space and open in the velocity space.
 The time advance of the system is implemented as follows;
(1) advection in the configuration space with a half time step $\Delta t/2$, (2) advance of the electromagnetic fields with a half time step, (3) acceleration by electric fields with a half time step, (4) gyro motion with a full time step $\Delta t$, (5) acceleration by electric fields with a half time step, (6) advection in the configuration space with a half time step, (7) current density calculation, (8) advance of the electromagnetic fields with a half time step. This algorithm is a second order leap-flog time integration, also employed in PIC simulations.

\subsubsection{Perpendicular wave propagation}\label{sec:perp-wave-prop}
We first test the wave propagation perpendicular to magnetic fields. The initial condition of the particle distribution function is given as follows,
\begin{eqnarray}
f_s(v_x,v_y,x,t=0) = \frac{n_s(x)}{\pi v_{s;th}^2} \exp\left[-\frac{v_x^2+v_y^2}{v_{s;th}^2}\right],\;\;\;\left(s=p,e\right),\label{eq:96}
\end{eqnarray}
where $v_{s;th}$ and $n_{s}(x)$ are the thermal velocity and the density distribution of $s$-species particles. We set a uniform density for both protons and electrons, and then add a small (1\%) uniform random noise only for the electron density. The initial magnetic field distribution is uniform, $E_y$ is zero, and $E_x$ is determined from the Poisson equation (\ref{eq:76}). Simulation parameters are $m_{p}/m_{e} = 16$, $v_{e;th} = 0.1$, $v_{p;th} = 0.025$ (corresponding to a ratio of the proton to electron temperature being unity), and a ratio of the electron gyro to plasma frequency $\omega_{ge}/\omega_{pe} = 0.5$. The simulation domain in the velocity space is $[-4v_{s;th},4v_{s;th}]$ with 50 grid points in both the $v_x$ and $v_y$ directions. The position is normalized by $\lambda_{D}$. The grid size in the configuration space is equal to $\lambda_{D}$, and the spatial length is $512\lambda_{D}$. The time step is $0.05/\sqrt{2}$.

\begin{figure}[htbp]
\centering
\includegraphics[clip,angle=0,scale=.30]{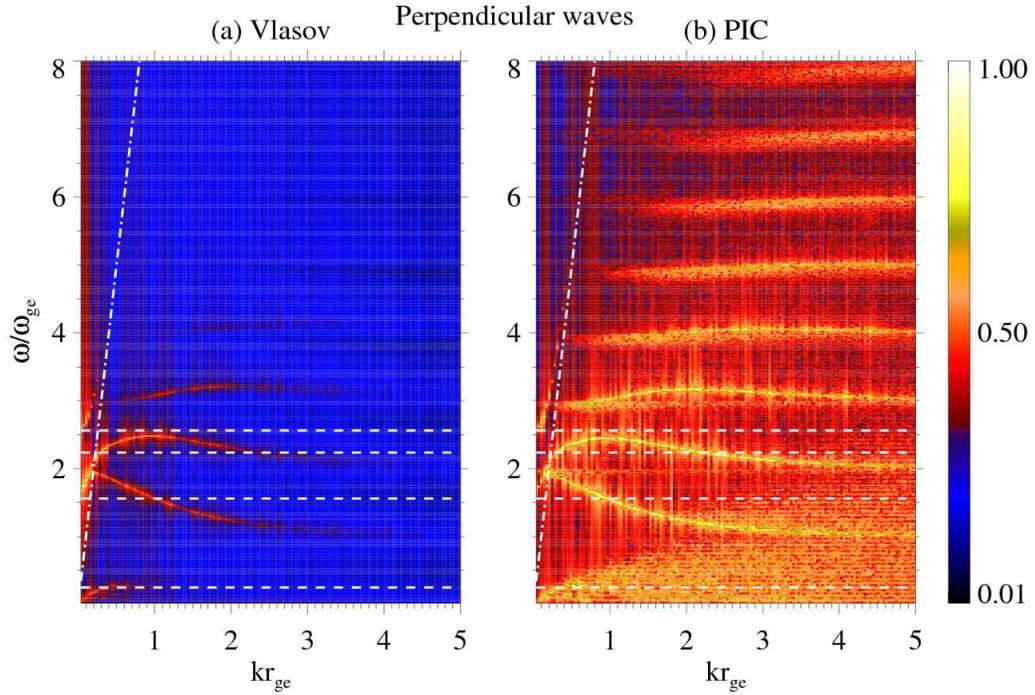}
\caption{Fourier spectra of perpendicular-propagating waves obtained from the electromagnetic (a) Vlasov simulation and (b) PIC simulation until $\omega_{pe} t=723.4$. Horizontal and vertical axes are the wavenumber and frequency normalized by the inverse electron gyro radius and the electron gyro frequency. Color contour shows the Fourier component of $E_x$ (normalized by its maximum value) to the power of 0.15 (for illustration). From top to bottom, dashed lines represent the R-mode cutoff, upper hybrid, L-mode cutoff, and lower hybrid frequencies. Dot-dashed lines represent the dispersion relation of the light mode in vacuum.}
\label{fig:10}
\end{figure}

Figure \ref{fig:10}(a) shows the Fourier spectrum of $E_x$ obtained from the electromagnetic Vlasov simulation integrated until $t=723.4$. For comparison, we also perform the electromagnetic PIC simulation with the same parameters, which is shown in Figure \ref{fig:10}(b). The number of particles is 5,000 in each cell so that the total memory usage is comparable between the two simulations. 
We can clearly identify the electron cyclotron (Bernstein) modes. We also identify the high-frequency X-mode and Z-mode at very low wavenumbers and frequencies close to the R- and L-mode cutoff ($\omega = 2.56\omega_{ge}$ and $1.56\omega_{ge}$), respectively. As expected, the Vlasov simulation provides a result with less noise, because it is free from the statistical noise. 
% Note that the amplitude of high-wavenumber waves is quite different between the two simulations. One of the reasons is the difference of the initial condition of $E_{x}$ ($\propto k^{-1}$ for Vlasov, $\propto k^{0}$ for PIC).

\begin{figure}[htbp]
\centering
\includegraphics[clip,angle=0,scale=.7]{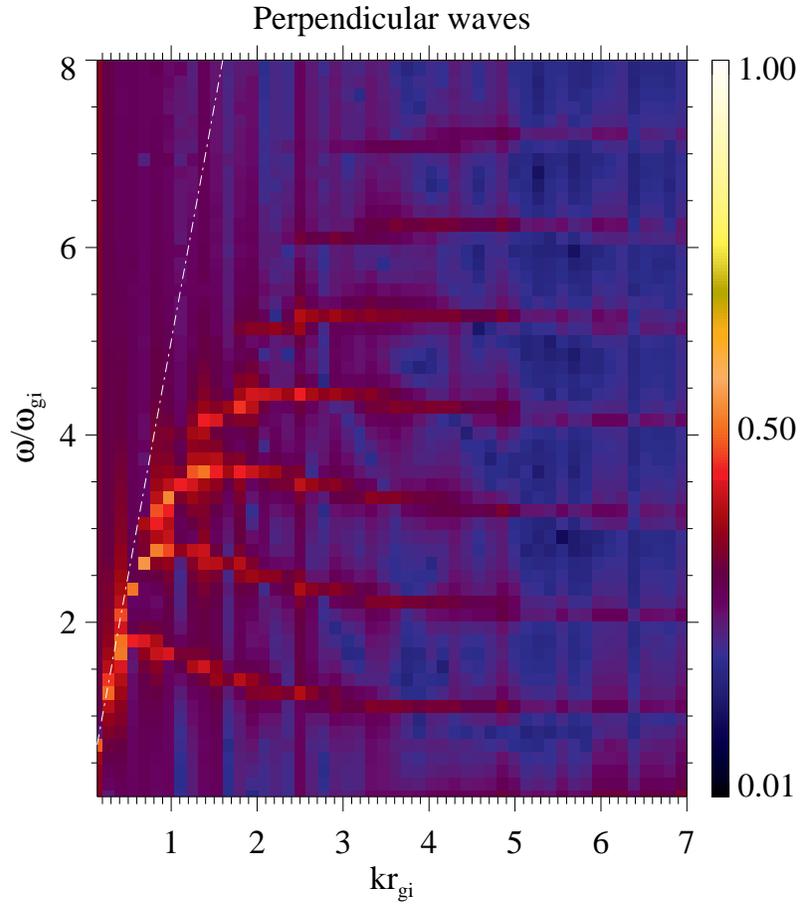}
\caption{Fourier spectrum of perpendicular-propagating waves obtained from the electromagnetic Vlasov simulation until $\omega_{pe} t=1447.4$. Horizontal and vertical axes are the wavenumber and frequency normalized by the inverse proton gyro radius and the proton gyro frequency. Color contour shows the Fourier component of $E_x$ (normalized by its maximum value) to the power of 0.15 (for illustration). A dot-dashed line represents the dispersion relation of the {\Alfven} wave.}
\label{fig:11}
\end{figure}

At very low wavenumber and frequency in Figure \ref{fig:10}(a), the contribution of protons is slightly seen. Then, we continue the same simulation for a long time until $t=1447.4$. 
% As a result, the ion Bernstein mode and the low-frequency X-mode (fast magnetosonic wave) are identified in Figure \ref{fig:11}. 
The ion Bernstein modes and the lower-hybrid waves can be seen in Figure \ref{fig:11}. 
Even at this moment (electrons gyrate more than a hundred times), the total energy is conserved very well within an error of $10^{-6}$, due to the fact that the MMA scheme can solve the solid body rotation problem with little numerical diffusion.

% Figure \ref{fig:12} also shows the Fourier spectrum obtained from the same simulation until $t=723.4$, but with a wider length $L_x = 512$. In addition to the Bernstein modes, branches of X- and Z-modes expected from the two fluid theory are seen at very low wavenumbers and frequencies close to the R- and L-mode cutoff, respectively.

\subsubsection{Harris current sheet equilibrium}\label{sec:harris-current-sheet}
We next test the one-dimensional Harris current sheet configuration, in which a dense plasma is confined at the center of the anti-parallel magnetic field configuration. The initial conditions of the particle distribution function and the magnetic field are given as a double Harris current sheet, for utilizing the periodic condition in the configuration space,
\begin{eqnarray}
f_s(v_x,v_y,x,t=0) &=& \frac{n_s^{-}(x)}{\pi v_{s;th}^2} \exp\left[-\frac{v_x^2+\left(v_y-u_{s}^{-}\right)^2}{v_{s;th}^2}\right]\nonumber \\
&+&\frac{n_s^{+}(x)}{\pi v_{s;th}^2} \exp\left[-\frac{v_x^2+\left(v_y-u_{s}^{+}\right)^2}{v_{s;th}^2}\right],\;\;\;\left(s=p,e\right),\label{eq:98}\\
n_{s}^{\pm}(x) &=& n_{0} {\rm sech}^2 \left(\frac{x \mp L/4}{\lambda}\right),\label{eq:99}\\
u_{s}^{\pm} &=& \pm \sgn\left(q_s\right) \frac{v_{s;th}^2}{\omega_{gs} \lambda},\label{eq:100}\\
B_{z}(x) &=& B_0 \left[\tanh \left(\frac{x+L/4}{\lambda}\right)-\tanh \left(\frac{x-L/4}{\lambda}\right)-1\right],\label{eq:101}
\end{eqnarray}
where $u_{s}^{\pm}$ is the diamagnetic drift velocity, $\omega_{gs} = |q_{s}|B_0/m_{s}c$ is the gyro frequency outside the sheet, $\lambda$ is the half thickness of the current sheet, and $L$ is the spatial length.
%  We do not superimpose an ambient uniform plasma that is often used for magnetic reconnection simulations \citep[e.g.,][]{2005PhRvL..94i5001D}. 
% The initial electric field is zero. 
The one-dimensional Harris current sheet is a steady-state solution of the Vlasov-Maxwell system, in which the plasma and magnetic pressures balance with each other. Therefore, it is a suitable benchmark test for electromagnetic Vlasov simulations: An accurate scheme is required for solving the advection and rotation in the presence of spatially-inhomogeneous plasma and magnetic field distributions and thereby keeping the equilibrium.

Simulation parameters are $m_p/m_e = 10$, $\omega_{ge}/\omega_{pe} = 0.1$, $v_{e;th}=0.0707$, $v_{p;th}=0.0224$ (corresponding to a temperature ratio of unity), $\lambda = d_{p} = c/\omega_{pp}$ (proton inertia length), and $L=1024 \lambda_{D}$.
The thermal velocity is determined so as to satisfy the pressure balance. The position is normalized by $d_{e}$. The simulation domain is $[-5v_{s;th},5v_{s;th}]$ in the velocity space with 32 grid points in both the $v_x$ and $v_y$ directions, and $[-L/2,L/2]$ in the configuration space with 1024 $(\Delta x = \lambda_{D})$ or 256 $(\Delta x = 4\lambda_{D})$ grid points. The time step is $0.02\pi$.

\begin{figure}[htbp]
\centering
\includegraphics[clip,angle=0,scale=.6]{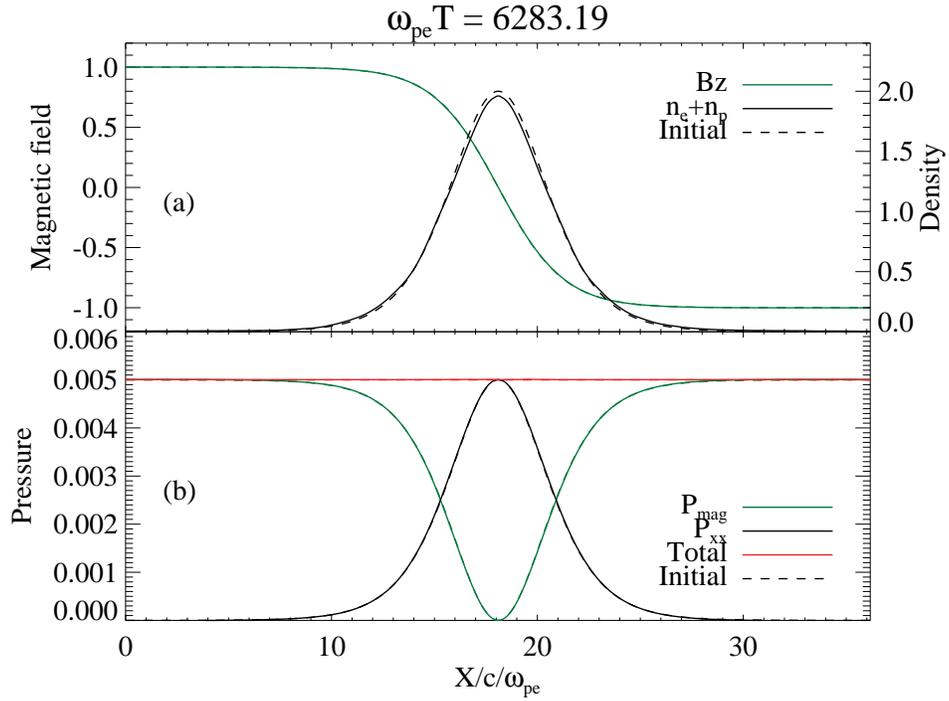}
\caption{One-dimensional Harris current sheet simulation with the MMA scheme. (a) The magnetic field (green lines) and density (black lines) distributions. (b) The pressure distribution. Green, black, and red lines are the magnetic, plasma, and total pressures, respectively. Solid lines are the simulation result and dashed lines are the initial. The grid size is $\lambda_{D}$.}
\label{fig:13}
\end{figure}

Figure \ref{fig:13}(a) shows the magnetic field and density $(n_e+n_p)$ distributions at $t=6283.19$ (corresponding to 100 gyrations of electrons outside the sheet), obtained from the simulation with $\Delta x = \lambda_{D}$. The simulation keeps the equilibrium of the current sheet with small numerical dissipation. The amplitude of numerically-produced electric fields is smaller than $10^{-3}$.

We check the pressure balance along the $x$ direction between the magnetic field and plasma,
\begin{eqnarray}
&& P_{\rm mag}(x) = \frac{B_{z}^2}{8 \pi},\nonumber \\
&& P_{xx}(x) = n_e T_{e;xx} + n_p T_{p;xx},\nonumber
\end{eqnarray}
% where $T_{s;xx}$ is the temperature in the $x$ direction. 
where $P_{xx}$ and $T_{s;xx}$ are the $x$-diagonal component of the pressure and temperature tensors.
In the simulation with the MMA scheme, diagonal components of the pressure tensor can be easily estimated from dependent variables,
\begin{eqnarray}
n T_{xx} &=& \inty{}{}\!\!\!\inty{}{}{} \left(v_x-V_{xb}\right)^2 fdv_xdv_y\nonumber \\
&=& \sum_{i,j}\left[2M_{v_x;i+1/2,j+1/2}^2-2V_{xb} M_{v_x;i+1/2,j+1/2}^1 + V_{xb}^2M_{i+1/2,j+1/2}^0\right],\nonumber
\end{eqnarray}  
where $V_{xb}$ is the bulk velocity in the $x$ direction, also estimated from them,
\begin{eqnarray}
V_{xb} = \frac{\inty{}{}\!\!\!\inty{}{}v_xfdv_xdv_y}{\inty{}{}\!\!\!\inty{}{}fdv_xdv_y} = \frac{\sum_{i,j}M_{v_x;i+1/2,j+1/2}^1}{\sum_{i,j}M_{i+1/2,j+1/2}^0}.\nonumber
\end{eqnarray}
% Since the MMA advance the zeroth to second order moments based on their governing equations, it is expected that the above macroscopic variables (pressure, bulk velocity, and also density) are determined with high precision.
Figure \ref{fig:13}(b) shows the pressure distribution. The pressure balance is preserved very well within an error of 0.1 \%. The plasma pressure is kept correct within an error of 2 \%.
%  The error corresponds to the energy density of numerically-generated electric fields ($E_x \sim 10^{-3}$).

\begin{figure}[htbp]
\centering
\includegraphics[clip,angle=0,scale=.6]{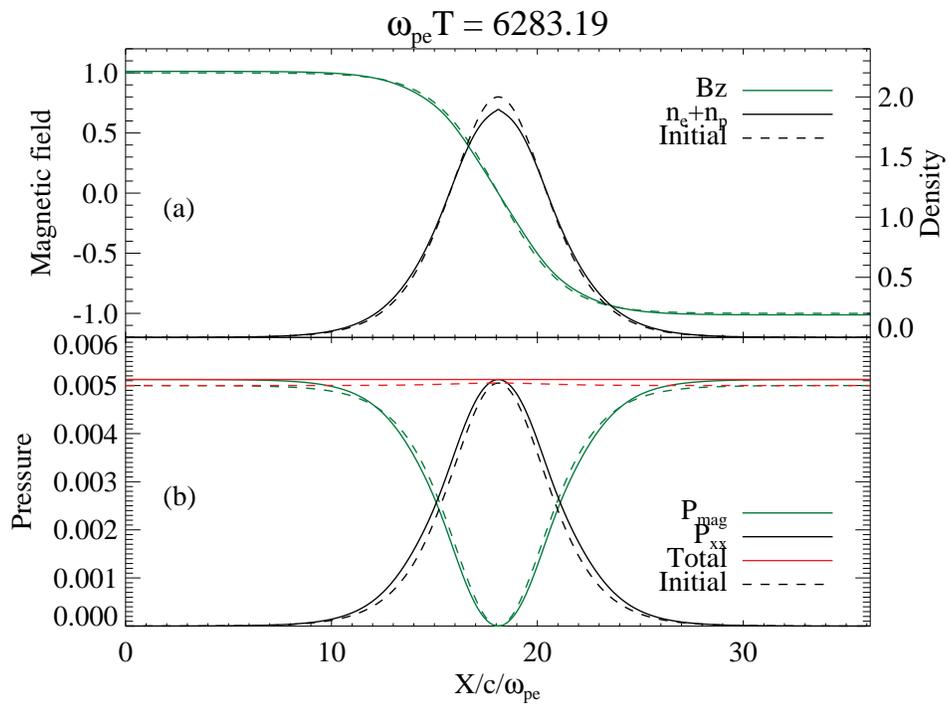}
\caption{The same as Figure \ref{fig:13}, but simulated with the CIP-CSL2 scheme. }
\label{fig:harris_csl}
\end{figure}

For comparison, we also perform the simulation in which the CIP-CSL2 scheme is applied to the advections in the velocity space as well as the configuration space. The number of grid points in the velocity space is $40\times40$ for this simulation so that the total memory usage is comparable to the simulation with the MMA scheme. The result is shown in Figure \ref{fig:harris_csl}. Compared to the MMA scheme, the CIP-CSL2 scheme causes the dissipation of the current sheet (increase of the sheet thickness), through the numerical heating of the plasma. The proton and electron pressures increase by 10 \% and 40 \% around the edge of the sheet, respectively. The total pressure increases by 2 \%.

\begin{figure}[htbp]
\centering
\includegraphics[clip,angle=0,scale=.6]{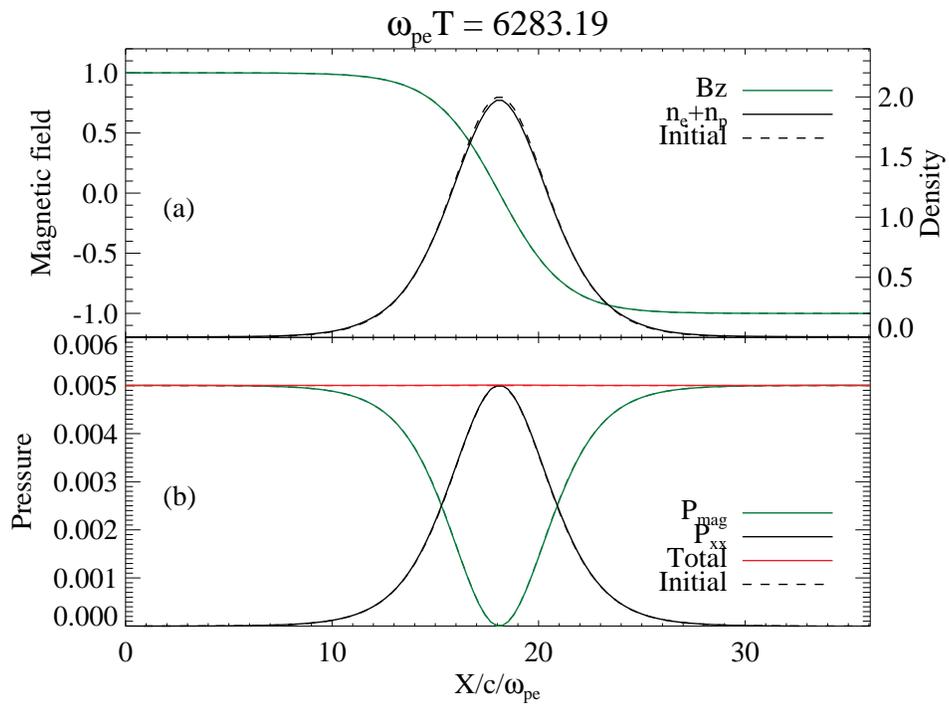}
\caption{The same as Figure \ref{fig:13}, but with the grid size $4\lambda_{D}$.}
\label{fig:14}
\end{figure}

Figure \ref{fig:14} shows the result from the same simulation with the MMA scheme, but with a larger grid size of $\Delta x = 4 \lambda_{D}$. 
The simulation again keeps the equilibrium of the current sheet.
The order of a numerical error is comparable to the result with $\Delta x = \lambda_{D}$.
% The result confirms that our electromagnetic Vlasov simulation successfully works even with the grid size larger than the Debye length, although a slight dissipation of the sheet is seen (top plot). 
The result confirms that our electromagnetic Vlasov simulation code is stable even with the grid size larger than the Debye length.
This is an advantage over an explicit PIC simulation, as suggested by, e.g., \cite{2009CoPhC.180..365U,2010PhPl...17e2311U}.

\section{Summary and Discussion}\label{sec:conclusion}
We have presented a new numerical scheme for solving the advection equation and the Vlasov equation.
The present scheme solves not only point values of a profile but also its zeroth to second order piecewise moments as dependent variables, for better conservation of the information entropy.
We have developed one- and two-dimensional schemes, and have shown their high capabilities.
% The scheme provides quite accurate solutions even with small numbers of grid points, although it needs to store a number of dependent variables on the memory.
The present scheme provides quite accurate solutions even with smaller numbers of grid points, although it requires a higher memory cost than other existing schemes.
% The MMA provides quite an accurate solution even with the small number of grids, although it treat a number of dependent variables. 
We have shown that, however, the total memory usage of the present scheme is smaller than the others for the same accuracy of solutions

% Applications of the one-dimensional scheme to the electrostatic Vlasov simulations (linear Landau damping, two stream instability, and beam propagation), and the two-dimensional scheme to the electromagnetic Vlasov simulations (perpendicular wave propagation, Harris current sheet equilibrium, and strictly perpendicular shock waves) have been presented.
Applications of the one-dimensional scheme to the electrostatic Vlasov simulations (linear Landau damping and two stream instability), and the two-dimensional scheme to the electromagnetic Vlasov simulations (perpendicular wave propagation and Harris current sheet equilibrium) have been presented.
%  Especially, the MMA2D allows us to perform the Vlasov simulation for a long time without the numerical diffusion. 
The two-dimensional scheme allows us to solve the gyro motion and the $\vect{E} \times \vect{B}$ drift motion for a long time with little numerical heating. 
Since the present scheme treats the zeroth to second order moments and advances them on the basis of their governing equations, the particle momentum and energy as well as mass are conserved very well. This is important for studying plasma phenomena such as convection, heating, and acceleration.
% Using the MMA, it will be possible to study the plasma convection, heating and acceleration processes based on the electromagnetic Vlasov simulation.

% The MMA performs a local interpolation of a profile. Therefore it is easy to implement the parallel computation. We consider that the hybrid parallel computation is useful for large-scale Vlasov simulations, in which each node accounts for a certain domain of the configuration space and the velocity space is allocated on a shared memory.
% A scheme for the advection in the configuration space is chosen so as to be parallelized in a distributed memory system.

Although the present scheme correctly solves the moments up to the second order, the entropy is numerically increased in the two stream instability (Section \ref{sec:two-stre-inst}), by the dissipation of fine structures in velocity space. 
This is understood that the perturbation of lower order moments decreases and higher order moments contribute to the entropy when the so-called filamentation phenomena proceed \cite[e.g.,][]{2005CoPhC.166...81P}. Since the filamentation is inevitable as long as discretizing velocity space, it is essentially impossible to exactly conserve the entropy with finite information.

The present scheme is designed specifically to solve the advection and rotation in velocity space in the Vlasov equation. In our Vlasov simulation code, the CIP-CSL2 scheme has been employed to solve the advection in configuration space. We note that, however, another scheme can be applied to the advection in configuration space and be combined with the present scheme.

Several advection schemes proposed for Vlasov simulations are designed to preserve positivity and non-oscillatory property \citep{2001JCoPh.172..166F,2008EP&S...60..773U}, in order to suppress a non-physical growth of plasma waves caused by numerically-produced positive gradient in velocity space. 
\cite{2006JPlPh..72.1057U} argued that the positivity-preserving and non-oscillatory are important properties for reliable Vlasov simulations.
% On the other hand, \cite{2002JCoPh.180..339A} have pointed out that dissipating fine structures is more important than preserving the positivity. 
% On the other hand, these schemes are relatively diffusive and may cause a non-physical global evolution driven by a plasma pressure gradient that is increased by the numerical diffusion in the velocity space, even when the plasma should be in an equilibrium state. This is serious especially for the electromagnetic Vlasov simulation. 
On the other hand, it may cause a non-physical global evolution driven by plasma pressure gradient that is increased by the numerical diffusion in velocity space, even when the plasma should be in an equilibrium state.
%  This is serious especially for the Vlasov simulation of magnetized plasmas, because existing schemes cause rapid numerical diffusion in the solid body rotation problem. 
Although the present scheme is not positivity-preserving or non-oscillatory, we have shown that the obtained results are better than the others. We thus consider that preserving high order moments is another important property for Vlasov simulations.  

Magnetized plasma phenomena that allow to assume the two dimensionality in velocity space are limited, e.g., strictly perpendicular shocks \citep{2009ApJ...690..244A} and the two-dimensional Kelvin-Helmholtz instability \citep{2006JGRA..11105213M,2010PhPl...17e2311U}. Even considering the one dimension in configuration space, there are many phenomena that should treat the full three-dimensional velocity space. 
We are now developing a three-dimensional scheme and its application to the full electromagnetic Vlasov simulation, which will enable us to study a wide variety of collisionless plasma phenomena. 
% This will be presented in near future. 

\section*{Acknowledgements}\label{sec:acknow}
We would like to thank T. Umeda, T. Miyoshi, T. Sugiyama, and K. Kusano for insightful comments on our manuscript, and anonymous referees for carefully reviewing the manuscript. T. M. is supported by a Grand-in-Aid for Young Scientists, (B) \#21740135.

\appendix
\section{Coefficients of the interpolation function of MMA2D}\label{sec:coefficients-mma2d}
{\footnotesize
\begin{eqnarray}
C_{11;i,j} &=& f_{i,j},\label{eq:45}\\
C_{12;i,j} &=& \frac{-1}{\Delta x} \left[2f_{i,j}+f_{iup,j}-\frac{9\sgn\left(\zeta_{i,j}\right)\sgn\left(\eta_{i,j}\right)}{\Delta x \Delta y^3} g(y_{j},y_{jup},M_{y;icell,jcell}^m)\right],\label{eq:46}\\
C_{13;i,j} &=& \frac{1}{\Delta x^2} \left[f_{i,j}+f_{iup,j}-\frac{6\sgn\left(\zeta_{i,j}\right)\sgn\left(\eta_{i,j}\right)}{\Delta x \Delta y^3} g(y_{j},y_{jup},M_{y;icell,jcell}^m)\right],\label{eq:47}\\
C_{21;i,j} &=& \frac{-1}{\Delta y} \left[2f_{i,j}+f_{i,jup}-\frac{9\sgn\left(\zeta_{i,j}\right)\sgn\left(\eta_{i,j}\right)}{\Delta x^3 \Delta y} g(x_{i},x_{iup},M_{x;icell,jcell}^m)\right],\label{eq:48}\\
C_{22;i,j} &=& \frac{1}{\Delta x \Delta y}\left[4f_{i,j}+2\left(f_{iup,j}+f_{i,jup}\right)+f_{iup,jup} - \frac{9\sgn \left(\zeta_{i,j}\right) \sgn \left(\eta_{i,j}\right)}{\Delta x \Delta y}\times \right.\nonumber \\
&& \left. \left\{\left(30\left\{\frac{x_i x_{iup}}{\Delta x^2}+\frac{y_j y_{jup}}{\Delta y^2}\right\}+2\left\{\frac{x_i}{\Delta x}+\frac{y_j}{\Delta y}\right\}+13\right)M^0_{icell,jcell} \right.\right.\nonumber \\
&& \left.\left. -4\left(\left\{8x_{iup}+7x_{i}\right\}\frac{M^1_{x;icell,jcell}}{\Delta x^2} + \left\{8y_{jup}+7y_{j}\right\}\frac{M^1_{y;icell,jcell}}{\Delta y^2}\right) \right.\right.\nonumber \\
&& \left. \left.+ 60\left(\frac{M^2_{x;icell,jcell}}{\Delta x^2} + \frac{M^2_{y;icell,jcell}}{\Delta y^2}\right)\right\}\right],\label{eq:49}\\
C_{23;i,j} &=& \frac{-1}{\Delta x^2 \Delta y}\left[2\left(f_{i,j}+f_{iup,j}\right)+f_{i,jup}+f_{iup,jup} - \frac{12\sgn \left(\zeta_{i,j}\right) \sgn \left(\eta_{i,j}\right)}{\Delta x \Delta y}\times \right.\nonumber \\
&& \left. \left\{\left(\frac{15 x_i x_{iup}}{\Delta x^2}+\frac{6y_{jup}^2+4y_{j}y_{jup}+5 y_j^2}{\Delta y^2}\right)M^0_{icell,jcell} \right.\right.\nonumber \\
&& \left.\left. -\left(15\left\{x_{iup}+x_{i}\right\}\frac{M^1_{x;icell,jcell}}{\Delta x^2} + 2\left\{8y_{jup}+7y_{j}\right\}\frac{M^1_{y;icell,jcell}}{\Delta y^2}\right)\right.\right.\nonumber \\
&& \left.\left. + 30\left(\frac{M^2_{x;icell,jcell}}{\Delta x^2} + \frac{M^2_{y;icell,jcell}}{\Delta y^2}\right)\right\}\right],\label{eq:50}\\
C_{31;i,j} &=& \frac{1}{\Delta y^2} \left[f_{i,j}+f_{i,jup}-\frac{6\sgn\left(\zeta_{i,j}\right)\sgn\left(\eta_{i,j}\right)}{\Delta x^3 \Delta y} g(x_{i},x_{iup},M_{x;icell,jcell}^m)\right],\label{eq:51}\\
C_{32;i,j} &=& \frac{-1}{\Delta x \Delta y^2}\left[2\left(f_{i,j}+f_{i,jup}\right)+f_{iup,j}+f_{iup,jup} - \frac{12\sgn \left(\zeta_{i,j}\right) \sgn \left(\eta_{i,j}\right)}{\Delta x \Delta y}\times \right.\nonumber \\
&& \left. \left\{\left(\frac{6x_{iup}^2+4x_{i}x_{iup}+5 x_i^2}{\Delta x^2}+\frac{15 y_j y_{jup}}{\Delta y^2}\right)M^0_{icell,jcell}\right.\right.\nonumber \\
&& \left.\left. -\left(2\left\{8x_{iup}+7x_{i}\right\}\frac{M^1_{x;icell,jcell}}{\Delta x^2} + 15\left\{y_{jup}+y_{j}\right\}\frac{M^1_{y;icell,jcell}}{\Delta y^2}\right)\right.\right.\nonumber \\
&& \left.\left. + 30\left(\frac{M^2_{x;icell,jcell}}{\Delta x^2} + \frac{M^2_{y;icell,jcell}}{\Delta y^2}\right)\right\}\right],\label{eq:52}\\
C_{33;i,j} &=& \frac{1}{\Delta x^2 \Delta y^2}\left[f_{i,j}+f_{iup,j}+f_{i,jup}+f_{iup,jup} - \frac{4\sgn \left(\zeta_{i,j}\right) \sgn \left(\eta_{i,j}\right)}{\Delta x \Delta y}\times \right.\nonumber \\
&& \left. \left\{\left(30\left\{\frac{x_i x_{iup}}{\Delta x^2}+\frac{y_j y_{jup}}{\Delta y^2}\right\}+11\right)M^0_{icell,jcell} \right.\right.\nonumber \\
&& \left.\left. -30\left(\left\{x_{iup}+x_{i}\right\}\frac{M^1_{x;icell,jcell}}{\Delta x^2} + \left\{y_{jup}+y_{j}\right\}\frac{M^1_{y;icell,jcell}}{\Delta y^2}\right)\right.\right.\nonumber \\
&& \left.\left. + 60\left(\frac{M^2_{x;icell,jcell}}{\Delta x^2} + \frac{M^2_{y;icell,jcell}}{\Delta y^2}\right)\right\}\right],\label{eq:53}
\end{eqnarray}
}where $\Delta x = x_{iup}-x_{i}$, $\Delta y = y_{jup} - y_{j}$, and,
\begin{eqnarray}
g(x_{i},x_{iup},M_{x;icell,jcell}^m) &=& \left(3x_{iup}^{2}+6x_{i}x_{iup}+x_{i}^2\right)M^0_{icell,jcell} \nonumber \\
&& -4\left(3x_{iup}+2x_{i}\right)M^1_{x;icell,jcell}\nonumber \\
&& +20M^2_{x;icell,jcell}.\nonumber
\end{eqnarray}

% Bibliography
\bibliographystyle{elsarticle-harv}
% \bibliography{ref}

% Figures
\clearpage
\gdef\thefigure{\arabic{figure}}

% \begin{figure}[htbp]
% \centering
% \includegraphics[clip,angle=0,scale=.5]{./f3.eps}
% \caption{Fourier spectra obtained from the one dimensional linear advection of a square wave with the MMA (left), CIP-CSL2 (center), and PFC (right) schemes. Horizontal and vertical axes are the wavenumber $k$ and frequency $\omega$ normalized by their Nyquist number. The exact dispersion relation is $\omega = k$. CFL = 0.1 is used.}
% \label{fig:3}
% \end{figure}

% \clearpage
% \begin{figure}[htbp]
% \centering
% \includegraphics[clip,angle=0,scale=.55]{./fbeam.eps}
% \caption{Electron phase space distributions (negative images with contours) at (a) $t=0.3$ and (b) $t=5.0$ in the beam propagation simulation. The time, position, and velocity are normalized by the inverse electron plasma frequency, Debye length, and thermal velocity, respectively.}
% \label{fig:beamprop}
% \end{figure}

% \begin{figure}[htbp]
% \centering
% \includegraphics[clip,angle=0,scale=.55]{./f7b.eps}
% \caption{The same as Figure \ref{fig:7}, but adjusts time steps so as to uniform the CPU time between the three simulations.}
% \label{fig:7b}
% \end{figure}

% \begin{figure}[htbp]
% \centering
% \includegraphics[clip,angle=0,scale=.7]{./fd.eps}
% \caption{Fourier spectrum of perpendicular-propagating waves obtained from the electromagnetic Vlasov simulation with a spatial length of 512$\lambda_{D}$. Horizontal and vertical axes are the wavenumber and frequency normalized by the inverse electron gyro radius and the electron gyro frequency. From top to bottom, dashed lines represent the R-mode cutoff, upper hybrid, L-mode cutoff, and lower hybrid frequencies.}
% \label{fig:12}
% \end{figure}

% \clearpage
% \begin{figure}[htbp]
% \centering
% \includegraphics[clip,angle=0,scale=.5]{./fshock1.eps}
% \caption{One-dimensional perpendicular shock simulation at $\omega_{pe} t = 3770$. (a,b) The electron phase space distributions in $(x,v_{ex})$ and $(x,v_{ey})$. (c,d) The proton phase space distributions in $(x,v_{px})$ and $(x,v_{py})$. (e,f,g) The electromagnetic field $(B_{z},E_{x},E_{y})$ distributions. The grid size is $\lambda_{D}$. The velocity and the electric field are normalized by the bulk flow velocity $u(0)$ and the motional electric field $E_{y0}$ at the upstream, respectively.}
% \label{fig:shock1}
% \end{figure}

% \clearpage
% \begin{figure}[htbp]
% \centering
% \includegraphics[clip,angle=0,scale=.6]{./fstack.eps}
% \caption{A stack plot of the magnetic field $B_{z}$ in the one-dimensional perpencidular shock simulation with the grid size $\lambda_{D}$.}
% \label{fig:fstack}
% \end{figure}

% \clearpage
% \begin{figure}[htbp]
% \centering
% \includegraphics[clip,angle=0,scale=.5]{./fshock2.eps}
% \caption{The same as Figure \ref{fig:shock1}, but with the grid size $4\lambda_{D}$.}
% \label{fig:shock2}
% \end{figure}

\end{document}